\documentclass[a4paper,fleqn,usenatbib,useAMS]{mnras}
\usepackage{newtxtext,newtxmath}
\usepackage{mathptmx}
\usepackage[T1]{fontenc}
\usepackage{amsmath}
\usepackage{amssymb}
\usepackage{graphicx}

\newcommand{\mh}{m_{\rm H}}
\newcommand{\nh}{n_{\rm H}}

\newcommand{\nth}{n_{\rm th}}
\newcommand{\cc}{{\rm cm^{-3}}}
\newcommand{\cs}{c_{\rm s}}
\newcommand{\vrot}{v_{\rm rot}}
\newcommand{\vkep}{v_{\rm Kep}}
\newcommand{\kms}{{\rm km~sec^{-1}}}
\newcommand{\lsun}{{\rm L}_\odot}
\newcommand{\rsun}{{\rm R}_\odot}
\newcommand{\msun}{{\rm M}_\odot}
\newcommand{\msunyr}{{\rm M}_\odot~{\rm yr}^{-1}}

\title[Fate of Pop~III multiples]{Formation and survival of Population III stellar systems}

\author[Hirano \& Bromm]{
Shingo~Hirano$^1$\thanks{E-mail: shirano@astro.as.utexas.edu}
and
Volker~Bromm$^1$ \\
$^1$Department of Astronomy, University of Texas, Austin, TX 78712, USA
}

\date{Accepted XXX. Received YYY; in original form ZZZ}
\pubyear{2017}

\begin{document}
\label{firstpage}
\pagerange{\pageref{firstpage}--\pageref{lastpage}}
\maketitle

\begin{abstract}
The initial mass function of the first, Population~III (Pop~III), stars plays a vital role in shaping galaxy formation and evolution in the early Universe.
One key remaining issue is the final fate of secondary protostars formed in the accretion disc, specifically whether they merge or survive.
We perform a suite of hydrodynamic simulations of the complex interplay between fragmentation, protostellar accretion, and merging inside dark matter minihaloes.
Instead of the traditional sink particle method, we employ a stiff equation of state approach, so that we can more robustly ascertain the viscous transport inside the disc.
The simulations show inside-out fragmentation because the gas collapses faster in the central region.
Fragments migrate on the viscous timescale, over which angular momentum is lost, enabling them to move towards the disc centre, where merging with the primary protostar can occur.
This process depends on the fragmentation scale, such that there is a maximum scale of $(1 - 5) \times 10^4$~au, inside which fragments can migrate to the primary protostar.
Viscous transport is active until radiative feedback from the primary protostar destroys the accretion disc.
The final mass spectrum and multiplicity thus crucially depends on the effect of viscosity in the disc.
The entire disc is subjected to efficient viscous transport in the primordial case with viscous parameter $\alpha \le 1$. 
An important aspect of this question is the survival probability of Pop~III binary systems, possible gravitational wave sources to be probed with the Advanced LIGO detectors.
\end{abstract}

\begin{keywords}
methods: numerical -- stars: formation -- stars: Population III -- dark ages, reionization, first stars -- early Universe
\end{keywords}

\section{Introduction} \label{sec:intro}

The first stars, the so-called Population~III (Pop~III), mark the epoch of cosmic dawn
by initiating reionization and the enrichment of the intergalactic medium (IGM) with heavy chemical elements \citep[e.g.][]{bromm11}.
The legacy of the first stars thus imprints the conditions for the subsequent
emergence of ordinary stellar populations, as detected in deep imaging with the {\it Hubble
Space Telescope} and large ground-based telescopes \citep[e.g.][]{finkelstein16}.
Understanding the formation of Pop~III stars is, therefore, crucial to elucidate the early history of the Universe.
The conditions of their formation out of primordial gas with no metals or dust
were significantly different from the present-day case. The wealth of observations
available to guide understanding of the latter may thus not be directly applicable to
the first star case. Given the highly complex nature of the star formation process,
theorists have performed numerical simulations to reveal the properties of the first
stars \citep[see][for recent reviews]{bromm13,greif15}.
Fortunately, the initial conditions in the Pop~III case are set by modern cosmology,
providing the successful $\Lambda$CDM model of structure formation, calibrated to
high precision by {\it WMAP} and {\it Planck} \citep[e.g.][]{PLANCK15XIII}.
Numerical simulations of first star formation
can thus follow an {\it ab-initio}, ``first principle'' based, approach.

Stellar evolution and death, which regulate the dynamical, radiative, and chemical feedback
on the surrounding medium, are largely determined by the stellar mass \citep[e.g.][]{schaerer02, heger02, yoon12, chatzopoulos12}.
Knowledge of the characteristic mass of the first stars, or their initial mass function (IMF), is thus
essential for understanding the early evolution of the first galaxies.
Because of the different thermal evolution during the gravitational collapse of a primordial cloud,
typical masses of the first stars are inferred to be higher than for more recent (Population~I and II) star formation \citep{abel02,bromm02}.
Numerous investigations have tried to determine the Pop~III IMF by performing numerical simulations, while fully taking into account the cosmological initial conditions.
Recently, such studies have begun to directly calculate the build-up
of the mass spectrum, extending into the radiation-hydrodynamical regime
when the gas accretion is terminated by protostellar radiative feedback.
The resultant stellar masses are a few tens to a few hundreds of solar masses
 \citep{hosokawa11, hosokawa12b, hosokawa16, stacy12, stacy16, susa13, susa14, hirano14}.
To understand the differences in the derived masses across the various studies,
one should compare numerical settings such as resolution, sub-grid physics,
and time coverage \citep[e.g.][]{stacy16}.

One important issue is the multiplicity of primordial star formation.
Fragmentation can occur on different scales during the star-formation process.
\cite{turk09} and \cite{clark11a} have shown that a primordial gas cloud
can fragment into multiple clumps in response to the rotation and turbulence
in the cloud, even before reaching protostellar scales. Once such
scales are reached, fragmentation in the protostellar disc becomes ubiquitous
\citep{stacy10, clark11b}.
This result was already anticipated by \cite{machida08} and \cite{saigo08} who have investigated the instability of a rotating protostellar core with a suite of parameterized simulations.
This could lead to the formation of small groups of multiple stars rather than a single star in each minihalo.
In this case, the available gas in the envelope is divided among multiple accretors, so that
the stars in such a multiple system could have relatively lower masses than in the case of a single star \citep[e.g.][]{peters10, susa14}.
However, there is an opposite effect, as well. A large fraction of the protostars could rapidly migrate inward due to efficient
gravitational torques, resulting in frequent mergers at the cloud centre \citep[e.g.][]{greif12,vorobyov13}.
Formation of massive stars could thus be enhanced \citep[e.g.][]{hosokawa16, sakurai16}.

Because the first stars are formed with a few tens of solar masses, the intriguing possibility arises that Pop~III binaries may leave massive black-hole (BH) binaries behind. 
Such massive BH binaries have recently been directly detected by the Advanced Laser Interferometer Gravitational Wave Observatory (aLIGO), challenging theorists to account for them \citep{GW150914, GW151226}. 
Among the possible candidates are Pop~III BH binaries \citep[e.g.][]{kinugawa14, hartwig16}. 
However, the key question then is whether the Pop~III pathway can realistically provide the close binaries required. 
We therefore need to verify the survivability of fragments during the formation of Pop~III companion stars.
This assessment requires simulations that simultaneously achieve high resolution, and follow the evolution over sufficiently long timescales. 
Such fully realistic simulations are extremely expensive computationally. 
Indeed, employing virtually no sub-grid prescriptions at all, \cite{greif12} could simulate only the first 10~yr after initial protostar formation. 
Traditionally, to avoid the computational impasse, numerical simulations use sink-particle techniques \citep{bate95} to study the long-term evolution of mass accretion and protostellar dynamics.
For the Pop~III case, such simulations have shown that the protostellar discs become gravitationally unstable and vigorously fragment \citep[e.g.][]{stacy10, clark11b, greif11, greif12, stacy13b,machida13,stacy16}.
However, the final fate of binaries is uncertain within this framework, because the sink particles cannot robustly model the merging or survival of protostars, e.g. by not allowing for the tidally-induced decay of orbits.

Here, we perform a series of hydrodynamic simulations to study the fate of fragments on different scales in the primordial star-forming cloud.
For this purpose, we adopt a stiffened equation-of-state approach to model fragmentation instead of the sink-particle method.
The sink-particle ($N$-body) technique cannot robustly model the dissipative encounter which is important for the energy and angular momentum transfer during the formation of close binary systems.
The remainder of the paper is organized as follows.
In Section~\ref{sec:the}, we summarize the key physics of the formation and evolution of Pop~III multiples.
In Section~\ref{sec:sim}, we describe our numerical methodology, followed by an exposition of the results (Section~\ref{sec:sim_res}).
Section~\ref{sec:dis} discusses the critical scale for the survival of the primordial gas fragments.
We finally offer concluding remarks in Section \ref{sec:sum}.

\section{Basic physical processes} \label{sec:the}

In this section, we summarize the physical processes for the formation and evolution of binary and small multiple stellar systems.
The elements to consider in primordial star formation are basically the same as in the present-day case.
The key difference is the presence or absence of metal and dust cooling, resulting in different thermal properties, such as the scale of chemo-thermal instability.

\subsection{Fragmentation} \label{sec:the_modes}

We first consider the formation of fragments in the star-forming cloud.
Specifically, there are two separate modes: cloud fragmentation during the initial collapse phase and disc fragmentation during the subsequent accretion phase.

\subsubsection{Initial collapse phase} \label{sec:the_modes_cloud}

In the collapsing cloud, the gravitationally-unstable scale is given by the Jeans mass \citep{abel02}
\begin{eqnarray}
M_{\rm J} \approx 1000~\msun~\left( \frac{T}{200~{\rm K}} \right)^{3/2} \left( \frac{\nh}{10^4~\cc} \right)^{-1/2}~,
\label{eq:Mjeans}
\end{eqnarray}
where $T$ is the gas temperature and $\nh$ the number density, normalized to the typical values encountered in primordial clouds.
The thermal evolution of the contracting cloud is controlled by the balance between compressional heating and radiative cooling processes.
During the collapse, the temperature initially decreases with density, but eventually increases again at the so-called loitering point.
When the mass enclosed in the contracting clump for the first time exceeds the Jeans mass, gravitational instability is triggered and run-away collapse ensues.

For primordial gas, the Jeans mass is typically a few hundred solar masses (Equ.~\ref{eq:Mjeans}), corresponding to a Jeans length of a few parsec:
\begin{eqnarray}
L_{\rm J} &=& \left( \frac{3}{4 \pi} \frac{M_{\rm J}}{\mh \nh} \right)^{1/3} ~,\\
&\approx& 3~{\rm pc}~\left( \frac{T}{200~{\rm K}} \right)^{1/2} \left( \frac{\nh}{10^4~\cc} \right)^{-1/2}~,
\label{eq:Ljeans}
\end{eqnarray}
where the proton mass is $\mh$, expressing the gas density as $\rho = \mh \nh$ for simplicity.
In general, the dynamical and thermal properties of a collapsing cloud, and thus the resulting fragmentation scale, are different among clouds.
The primordial star-forming cloud is formed via gravitational compression by the host dark matter minihalo.
Cloud conditions are partially shaped by the turbulence that originates in the dark matter substructure, which differs among host haloes \citep{hummel15,hummel16}.
To constrain the extent of this cosmic variance, we analyze a number of clouds obtained from earlier {\it ab initio} cosmological simulations \citep{hirano14,hirano15a}.

\subsubsection{Accretion phase} \label{sec:the_modes_disk}

After initial protostar formation, the circumstellar disc increases in mass via gas accretion.
The disc is driven towards gravitational instability, resulting in ubiquitous multi-scale fragmentation throughout the disc, as shown in many previous studies \citep{clark11b,greif11,greif12,stacy13b,susa14,stacy16,hosokawa16}.
The smallest fragmentation scale is that of primary protostar formation, with a mass scale of 0.01~$\msun$ and a length scale of 0.03~au \citep{yoshida08}.
The fragmentation scale increases with increasing disc mass because the local Jeans mass becomes larger in the outer regions of the  disc.
We can understand this behaviour as follows.
The temperature increase after the loitering point can be approximated as
\begin{eqnarray}
T &\approx& 200~{\rm K}~\left( \frac{\nh}{10^{4}~\cc} \right)^{0.1}~,\\
&=& 2500~{\rm K}~\left( \frac{\nh}{10^{15}~\cc} \right)^{0.1}~,
\label{eq:Temp-Dens}
\end{eqnarray}
over the density range $10^4 < \nh / \cc <10^{16}$ \citep[see fig.~3 in][]{yoshida06}.
Substituting into the equation for the Jeans mass (Equ.~\ref{eq:Mjeans}), we find
\begin{eqnarray}
M_{\rm J} &\approx& 1000~\msun~\left( \frac{\nh}{10^4~\cc} \right)^{-0.35}~,\\
&=& 0.15~\msun~\left( \frac{\nh}{10^{15}~\cc} \right)^{-0.35}~.
\label{eq:Mjeans-Dens}
\end{eqnarray}
Thus, the fragment mass scale in the outer disc, where densities are lower, increases.

\subsection{Drivers of multiple evolution} \label{sec:the_bin}

The newborn stellar binary or multiple is unstable because dynamical and dissipative effects can easily break down the system.
We have to consider the time evolution of the system until it reaches a stable environment.
In particular, we will also discuss the possible formation paths of massive close binaries, given their importance.
Initially, the newborn multiple is embedded in a gas-rich environment, such that dissipative encounters can carry away the angular momentum of the system's components and drive their merger.
Furthermore, gravitational interaction within the multiple system can also bring the binary components closer together.
Merging will end when the mechanism of angular momentum transport, dynamical or dissipative, ceases to act.

\subsubsection{Physics of multiple survival} \label{sec:the_bin_survival}

To shut off dissipative merging, the gaseous material in the disc needs to be depleted via gas accretion and radiative feedback.
In the latter case, the important mechanism is the UV radiation feedback from the protostar which can photo-dissociate and sweep away the surrounding gas \citep{mckee08}.
\begin{figure}\begin{center}
\includegraphics[width=\columnwidth]{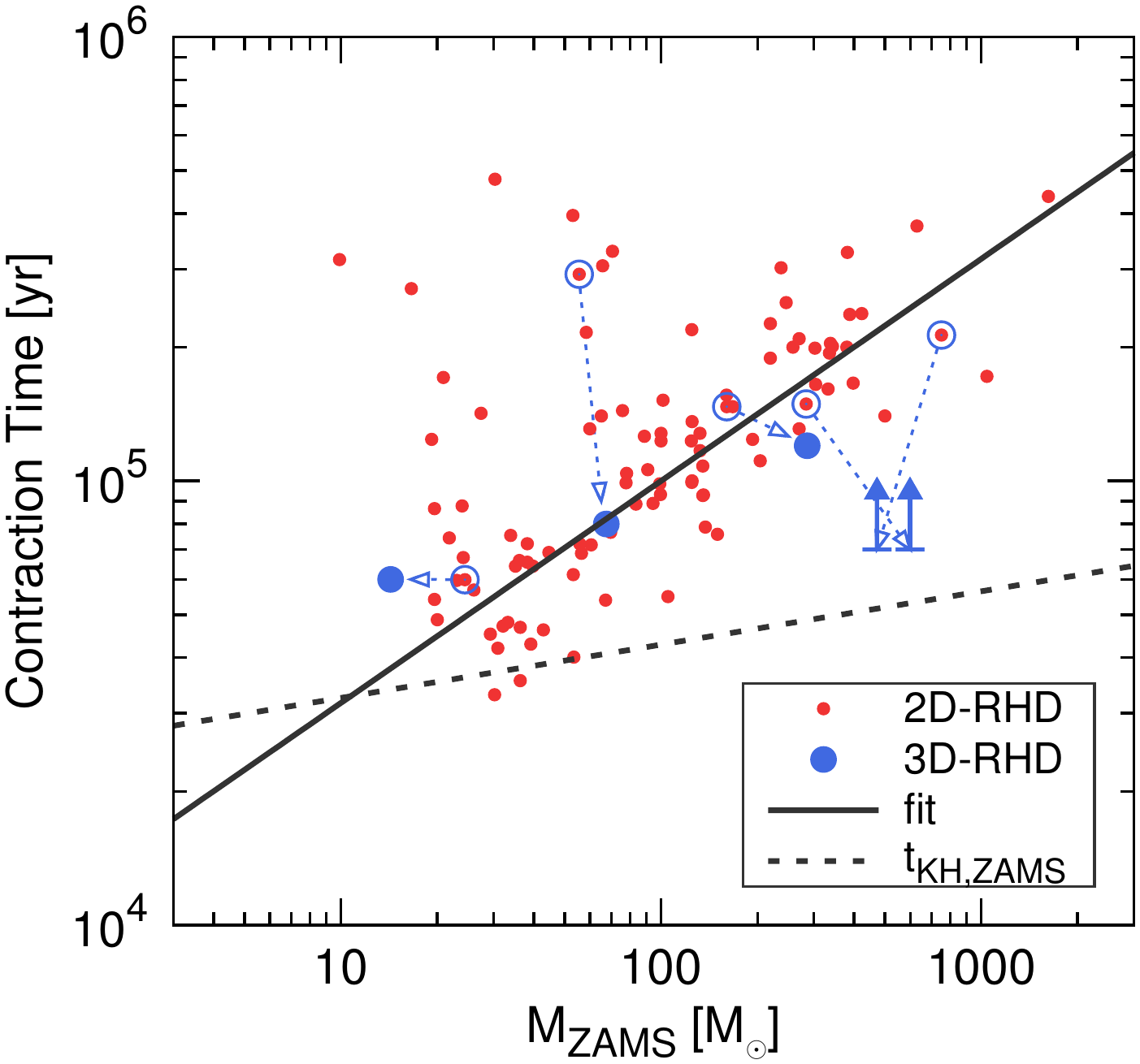}
\caption{
Contraction time during the protostellar accretion phase from core formation to the zero-age main-sequence (ZAMS) phase, as a function of the stellar mass reached on the ZAMS.
The red dots show two-dimensional (2D)-RHD simulation results \citep{hirano14}, whereas the larger blue circles mark three-dimensional (3D)-RHD calculations \citep{hosokawa16}.
The two arrows represent results from the latter class, but for cases where the accretion has not yet stopped.
The open circles and dotted arrows illustrate the difference between calculation results for the same initial conditions, but with different simulation methods (2D and 3D-RHD).
The solid line is a simple fitting function to reproduce the numerical data, where $t=10^5~{\rm yr}~(M_{\rm ZAMS}/100~\msun)^{1/2}$.
The dotted line shows the Kelvin-Helmholtz (KH) timescale for ZAMS stars (Equ.~\ref{eq:KHtime_ZAMS}).
}
\label{fig:Mstar-Tacc}
\end{center}\end{figure}
This feedback becomes effective once the star enters the main-sequence phase.
We thus need to consider the evolution for at least the Kelvin-Helmholtz (KH) timescale, over which the protostar radiates away its gravitational energy, defined as
\begin{eqnarray}
t_{\rm KH} \equiv \frac{G M_*^2}{R_* L_*}~,
\label{eq:KHtime}
\end{eqnarray}
where $G$ is Newton's constant, and $M_*$, $R_*$, and $L_*$, the stellar mass, radius, and luminosity, respectively.
The relations between these properties on the the zero-age main-sequence (ZAMS) are well fitted as
\begin{eqnarray}
R_{\rm ZAMS} &=& 4.61~\rsun~\left( \frac{M_*}{100~\msun} \right)^{0.58}~,\\
L_{\rm ZAMS} &=& 1.57 \times 10^6~\lsun~\left( \frac{M_*}{100~\msun} \right)^{1.30}~,
\label{eq:ZAMS}
\end{eqnarray}
\citep[see equ. B5 and B6 in][]{hirano14}.
With these expressions, the KH timescale becomes
\begin{eqnarray}
t_{\rm KH, ZAMS} = 4.28 \times 10^4~{\rm yr}~\left( \frac{M_*}{100~\msun} \right)^{0.12}~.
\label{eq:KHtime_ZAMS}
\end{eqnarray}
Some previous studies have carried out sophisticated calculations, tracing the evolution of the accreting protostar until the ZAMS is reached.
In Fig.~\ref{fig:Mstar-Tacc}, we compare results from such exact evolutionary calculations with $t_{\rm KH, ZAMS}$.
As can be seen, the dependence of contraction time on ZAMS mass is different from the simple KH scaling.
This is because KH contraction represents only part of protostar evolution, where contraction is delayed in cases of continued accretion.
For the target mass range in this study, stars with $10 < M_{\rm ZAMS}/\msun < 100$, the time to initiate strong radiative feedback is $3 \times 10^4$ to $10^5$~yr.

\subsubsection{Physics of binary mergers} \label{sec:the_bin_ms}

After reaching the main sequence, Pop~III stars evolve over a lifetime of a few Myr \citep{schaerer02}.
Finally, for masses of a few tens of solar masses, they end their life as massive black holes, possibly without triggering any explosion.
The conditions for a Pop~III BH binary system to merge within a Hubble time, to produce an observable gravitational wave signal, are as follows:

\paragraph*{Binary masses.}
As an example, the recently detected GW signal can be emitted by the coalescence of a BH binary with component masses of $\sim 30~\msun$ \citep{GW150914}.
Typical Pop~III stellar masses, determined by numerical simulations, are of similar value \citep[e.g.][]{hosokawa11,stacy12,susa14}.
Furthermore, a more massive primary star with $> 50~\msun$ evolves into a red supergiant, engulfing the secondary star in a large common envelope phase \citep{kinugawa14}.
The mass transfer between binary stars continues until a large fraction of the hydrogen envelope, $1/2 - 1/3$ of the initial mass, is ejected.
Thus, a Pop~III progenitor binary system with $> 50$ and $\sim 30~\msun$ can result in a BH binary with $\sim 30~\msun$ each.

\paragraph*{Binary separation.}
The merger timescale of a black hole binary is inversely proportional to the third power of mass and the forth power of separation,
\begin{eqnarray}
	t_{\rm coal} = 0.6~{\rm Gyr} \left( \frac{a_0}{0.1~{\rm au}} \right)^4 \left( \frac{M_1}{30~\msun} \frac{M_2}{30~\msun} \frac{M_1 + M_2}{60~\msun} \right)^{-1} \, ,
	\label{eq:t_coal}
\end{eqnarray}
where $a_0$ is the initial semi-major axis, and $M_1$ and $M_2$ are the masses of the primary and secondary stars \citep[equ. 80 in][]{kinugawa14}.
There is a strong constraint on the initial separation of about a few 0.1~au to allow coalescence within the Hubble time.
If the binary system were embedded within a diffuse gaseous component, close approach could readily occur due to viscous angular momentum transfer.\\

Therefore, suitable candidates for close, massive BH binaries that could give rise to a detectable GW signal should exist within the ensemble of Pop~III stellar systems with parameters ($M_1$, $M_2$, $a_0$).

\subsection{Formation of massive close binaries} \label{sec:the_bin_massiveclose}

Thus, we have to determine the masses and separation of Pop~III binaries at the stage when the surrounding gaseous material disappears.
In this context, a key quantity is the contraction time of the primary protostar (Fig.~\ref{fig:Mstar-Tacc}),
because the radiative feedback which sweeps away the surrounding gas, thus terminating the mass accretion, becomes effective only after the KH contraction to the ZAMS \citep{mckee08}.
Once the primary star reaches the ZAMS and photo-evaporates the surrounding medium, the possibility for dissipative interactions greatly decreases, such that the binary separation cannot easily decrease further.\\

As discussed above, there are multiple mechanisms that can influence the final fate of a Pop~III multiple system.
To examine this inherently complex problem, numerical simulations are essential.
In the remainder of the paper, we present our custom-designed hydrodynamical simulations to elucidate the impact of viscous processes on the long-term fate of Pop~III stellar systems.

\section{Numerical Methodology} \label{sec:sim}

To assess the final fate of Pop~III multiple systems, with a primary focus on binaries, we have to examine the long-term evolution of fragments until the strong radiative feedback from massive stars evaporates the natal cloud gas, comprising a time interval of about $10^5$~yr (see Fig.~\ref{fig:Mstar-Tacc}).
Fully realistic simulations which can directly resolve the newborn protostar, with initial mass of 0.01~$\msun$ and radius of 1~$\rsun$ \citep{yoshida08}, are computationally very expensive, such that it is currently prohibitive to continue  them through such an extended time \citep{greif12}.
Furthermore, judging the final fate of fragments that form farther out in the disc requires longer time scales, as well.
Therefore, we carry out three-dimensional hydrodynamic simulations with different resolutions to examine fragmentation inside the primordial star-forming cloud at different scales (Section~\ref{sec:sim_met_hydro}).
For this purpose, we employ a stiff equation of state approach, where the small-scale evolution is not resolved, but where fragments can viscously interact with the surrounding gaseous material, unlike the collisionless sink particles (Section~\ref{sec:met_frag}).

\begin{figure}\begin{center}
\includegraphics[width=\columnwidth]{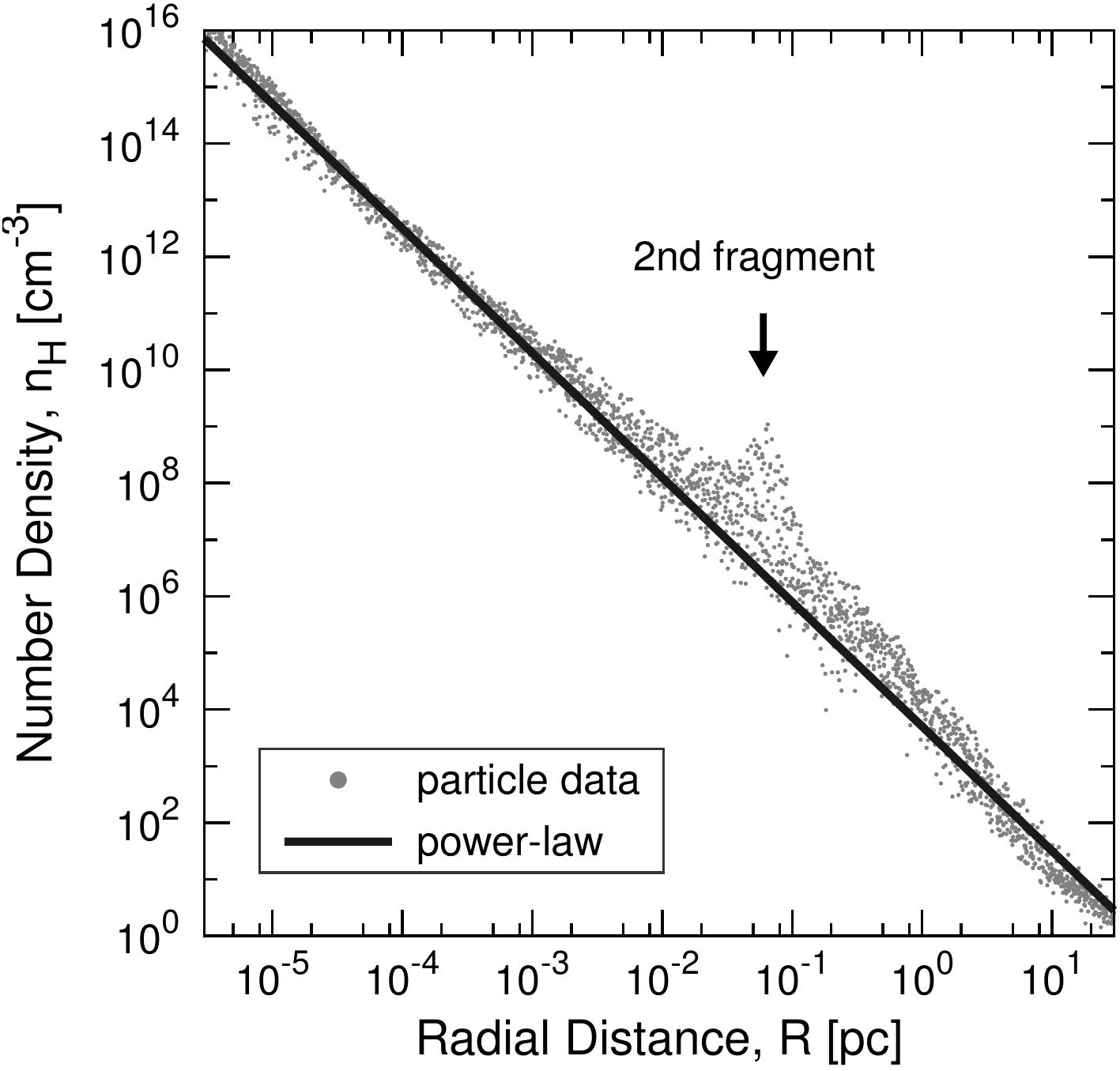}
\caption{
Radial gas density profile of the primordial star-forming cloud, shown for the final snapshot without imposing a density threshold (see Sec.~\ref{sec:sim_res_t0}).
The dots represent individual SPH particles.
During the collapse, secondary fragmentation occurs at about 0.05~pc from the primary density peak.
The solid line shows a power-law density distribution, where $\nh \simeq 5000~\cc~(R/{\rm pc})^{-2.2} = 2.5 \times 10^{15}~\cc~(R/{\rm au})^{-2.2}$ \citep[see][for the slope]{omukai98}.
}
\label{fig:Radi-Dens}
\end{center}\end{figure}

\subsection{Initial conditions} \label{sec:sim_met_init}

A series of hydrodynamic simulations is started from a primordial star-forming cloud found in the cosmological volume of \cite{hirano14}.
The gravitationally collapsing cloud has one widely separated pair of density peaks before primary protostar formation.
The cosmological simulation found 6 candidate clouds with secondary fragments \citep[see fig.~19 and table~5 in][]{hirano14}, and here we focus on one of them, which contains the closest fragment pair (see Fig.~\ref{fig:Radi-Dens}).
Below, we briefly summarize the procedure in carrying out this parent simulation.

The cosmological initial conditions were generated by using a modified version of {\sc n-genic} \citep{springel05}.
We adopted the standard $\Lambda$-Cold Dark Matter ($\Lambda$CDM) cosmology, with parameters \citep{komatsu11} as follows: a total matter density of $\Omega_{\rm m} = 0.272$, baryon density of $\Omega_{\rm b} = 0.0456$, dark energy density of $\Omega_{\Lambda} = 0.728$, all in units of the critical density, a Hubble constant of $h = 0.704$, and a spectral index of $n_{\rm s} = 0.963$.
We use the power spectrum given in \cite{eisenstein99}, normalized to $\sigma_8 = 0.809$.
The cosmological simulation was initialized at redshift $z_{\rm init} = 99$, employing periodic boundary conditions in a computational box with a linear size of $1~h^{-1}$~Mpc.
To achieve sufficient resolution for the primordial star-forming cloud with mass of $1000~\msun$ (Equ.~\ref{eq:Mjeans}), we performed a hierarchical zoom-in simulation. The resulting gas particle mass in the refined region is $3.49~\msun$, such that the target cloud can be resolved with more than 100 particles.

The three-dimensional hydrodynamic simulations were carried out with the parallel $N$-body/Smoothed Particle Hydrodynamics (SPH) code {\sc Gadget-2} \citep{springel05}, suitably modified for the primordial star formation case.
The chemical rate equations are solved for 14 primordial species (${\rm e^-}$, ${\rm H}$, ${\rm H^+}$, ${\rm H^-}$, ${\rm He}$, ${\rm He^+}$, ${\rm He^{++}}$, H$_2$, ${\rm H_2^+}$, ${\rm D}$, ${\rm D^+}$, ${\rm HD}$, ${\rm HD^+}$, and ${\rm HD^-}$), as in \cite{yoshida06,yoshida07}.
The code employs the Sobolev method for H$_2$ line cooling \citep{yoshida06}, and a ray-tracing approach for continuum cooling by H$_2$ collision-induced emission \citep[CIE, which becomes efficient at $\nh > 10^{14}~\cc$, see][]{yoshida08,hirano13}.
To achieve large dynamic range in the runaway collapsing cloud, the calculation also adopted the particle-splitting technique of \cite{kitsionas02}, with the refinement criterion that the local Jeans length is always resolved by 15 times the local smoothing length.
Finally, the gas particle mass had decreased to $\sim 10^{-5}~\msun$, resulting in a nominal mass resolution of $\sim 10^{-3}~\msun$.

For the case considered here, the simulation was terminated when the gas number density in the centre of the collapsing cloud reached $10^{13}~\cc$ at $z = 16.3$.
At this time, the host dark matter minihalo is characterized by a virial radius of $94$~pc, and a virial mass of $5.0 \times 10^5~\msun$.
The collapsing clump is located at the centre of a gravitationally unstable cloud whose Jeans length is $R_{\rm J} = 0.8~{\rm pc}$ and mass $M_{\rm J} = 1500~\msun$.
In Fig.~\ref{fig:Radi-Dens}, we show the gas density profile around the collapsing clump.
The density distribution is well reproduced with a power-law of $\propto R^{-2.2}$, but there is another density peak, located about $0.05~{\rm pc} \simeq 10^4~{\rm au}$ from the primary clump.
Thus, these two clumps formed from the same gravitationally unstable parent cloud.

\subsection{Hydrodynamic simulations} \label{sec:sim_met_hydro}

The parent simulation continues during the cloud collapse phase and ends before the protostar formation (Section~\ref{sec:sim_met_init}).
Our goal is to examine the formation and long-term evolution of fragmentation on different scales during the  protostellar accretion phase.
For this purpose, we rerun a series of hydrodynamical simulations, employing our artificial optically-thick core methodology (Section~\ref{sec:met_frag}).

The re-runs are initiated from a snapshot of the parent simulation when the peak number density has reached $10^6~\cc$, before the second fragment forms.
The calculation settings are matched to the parent cosmological run (Section~\ref{sec:sim_met_init}).
To reduce the computational cost and examine the long-term evolution, we adopt a minimum refinement criterion where the Jeans mass is always resolved by more than 50 gas particles, $N_{\rm res} = M_{\rm J}/m_{\rm sph} \geq 50$, comparable to the size of the SPH kernel.
This resolution is less than that enforced in the parent simulation, where $N_{\rm res} \geq 10^4$, which is necessary to investigate the evolution of turbulence \citep[e.g.][]{turk09}. 
Our reduced resolution adopted here, however, is sufficient to examine the hydrodynamics of fragmentation, the main target of our study \citep{bate97}.
The cloud and disc evolution are followed after primary protostar formation.
We here neglect the effect of protostellar radiation feedback on the accreting gas to reduce the computational cost.
Once the radiation feedback becomes effective, photo-evaporating the surrounding medium, the migration process of fragments via angular momentum transport ends.
Our main interest here is the formation and evolution of fragments until such a terminal phase, thus justifying our neglect of radiative transfer effects.

\subsection{Modeling fragmentation without sinks} \label{sec:met_frag}

We introduce an opaque core model which artificially reduces radiative cooling for gas particles whose density exceeds a threshold value, $\nth$.
Specifically, we impose an artificial optical depth,
\begin{eqnarray}
\tau_{\rm art} = \left( \frac{\nh}{\nth} \right)^2~,
\label{eq:ModelOpacity}
\end{eqnarray}
depending only on the local number density of the gas, $\nh$.
We adopt this particular formulation with a square dependence on density to reproduce the canonical adiabatic compression law.
With the optical depth in hand, we can calculate an escape fraction of
\begin{eqnarray}
\beta_{\rm esc,art} = \frac{1 - \exp(-\tau_{\rm art})}{\tau_{\rm art}}~.
\label{eq:ModelEscape}
\end{eqnarray}
This modeling assumes a spherical and uniform density structure around SPH particles.
In our simulations, all radiative cooling rates $\Lambda$ are modified as follows
\begin{eqnarray}
\Lambda_{\rm red} = \beta_{\rm esc,art} \cdot \Lambda_{\rm thin}~.
\label{eq:ModelRadCoolRate}
\end{eqnarray}
Through artificially reduced cooling, dense regions exceeding the threshold density are thus experiencing compressional heating, leading to the formation of a hydrostatic core in the centre of a collapsing flow.
Therefore, the complicated hydrodynamics occurring inside the opaque core, which would require extreme computational cost, can be removed from the simulations.
The opaque cores thus created replace the sink particles used in the majority of current simulation work.

\begin{figure}\begin{center}
\includegraphics[width=\columnwidth]{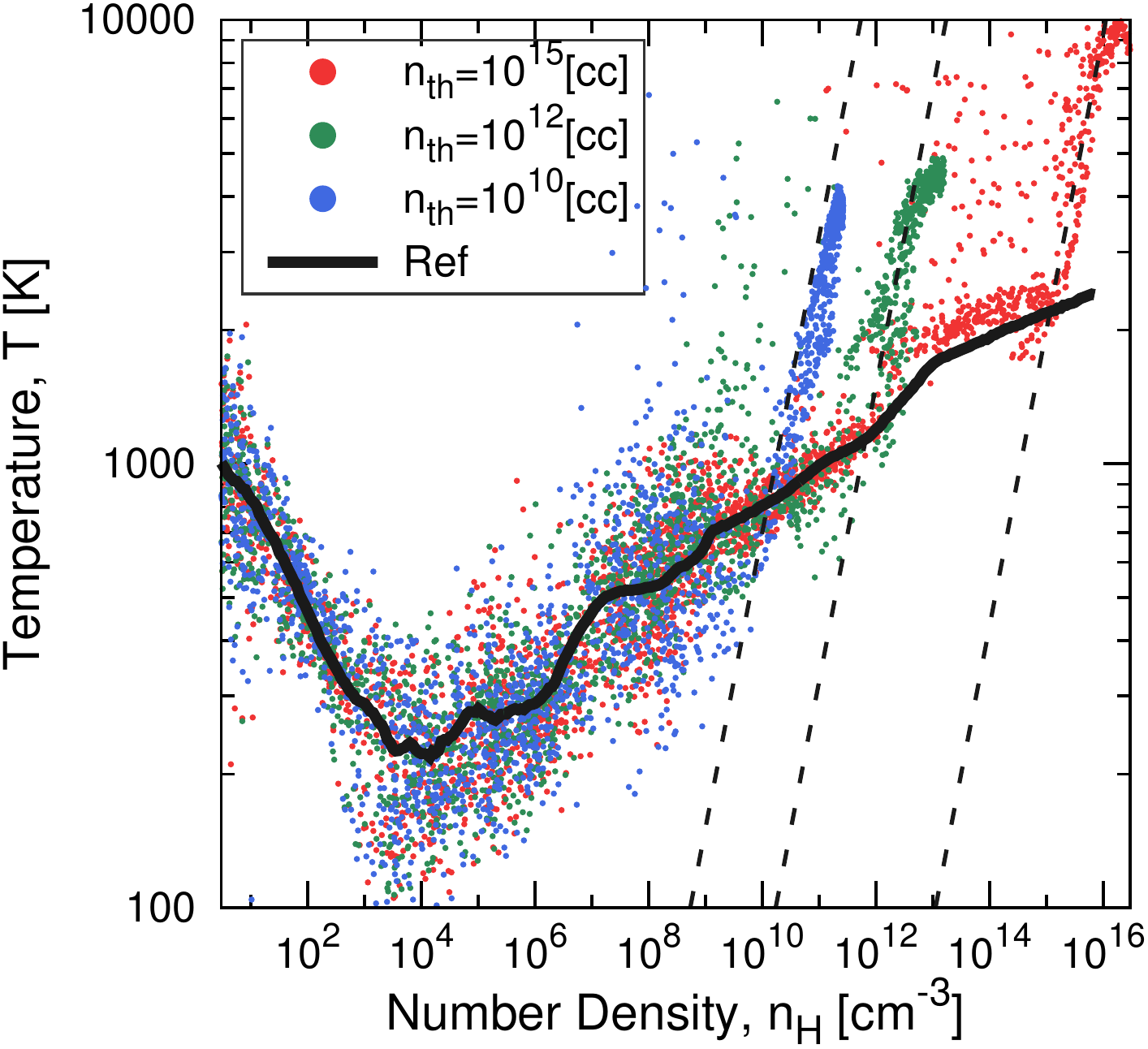}
\caption{
Gas temperature profiles of primordial star-forming cloud as a function of number density.
Coloured dots represent SPH particles for simulations with different threshold densities for the onset of artificial opacity, $\nth / \cc = 10^{15}$ (red; at $100$~yr after the central density reaches $\nth$), $10^{12}$ (green; at $10^3$~yr), and $10^{10}$ (blue; at $10^4$~yr), respectively.
The solid line shows the reference run without artificial opacity.
The three dashed lines illustrate the behaviour under adiabatic compression, $T \propto \nh^{\gamma-1}$ where $\gamma = 5/3$, nicely reproducing the simulated heat-up in the opaque cores.
}
\label{fig:Dens-Temp}
\end{center}\end{figure}

Within this model, we control the resolution scale by adjusting the threshold density.
To examine fragmentation on different scales, we perform simulations with three threshold densities, $\nth = 10^{10}, 10^{12}$, and $10^{15}~\cc$.
The corresponding resolution in mass is $M_{\rm res} = N_{\rm res} m_{\rm sph} \sim 0.1$, $0.1$, and $0.01~\msun$, and in length
$L_{\rm res} \simeq (3/4\pi \cdot M_{\rm res}/\mh\nth)^{1/3} \sim 34.5$, $7.43$, and $0.345$~au, respectively.\footnote{The former two cases have the same mass resolution but different length resolutions, because their target densities are different, $\nh = 10^{10}$ and $10^{12}~\cc$.}
Fig.~\ref{fig:Dens-Temp} illustrates the formation of a hot, opaque core by considering the evolution in the density-temperature plane.
Compared to a reference run without artificial opacity, the gas temperature quickly rises once the density has exceeded the threshold value. Specifically,
the resultant temperature slopes well reproduce the adiabatic compression law, $T \propto \nh^{\gamma - 1}$, with an adiabatic index of $\gamma = 5/3$, appropriate for monatomic gas.
Thus, our artificial optical depth model is able to stop the cloud collapse by reducing the radiative cooling rates.
The Jeans scale (Equ.~\ref{eq:Mjeans} and \ref{eq:Ljeans}) at the respective threshold densities and corresponding temperatures, $T \sim 900$, $1000$, and $2000$~K for $\nth = 10^{10}$, $10^{12}$, and $10^{15}~\cc$, resulting in $M_{\rm J} \sim 9.5$, $1.1$, and $0.1~\msun$ ($L_{\rm J} \sim 1300$, $140$, and $6.2$~au), is well resolved in our simulations.

A fundamental limitation of our methodology is that the viscous dissipation experienced by the puffed-up cores is artificially enhanced.
Similarly, the artificially reduced density concentration renders the envelope gas easier to strip away.
These limitations are not serious, as long as we focus on the dynamics on scales larger than the numerical resolution.
We discuss the caveats arising from using the artificial hydrostatic core model in greater detail in Section~\ref{sec:caveat}.

\begin{table*}
\begin{center}

\begin{tabular}{llrrrcrrr}
\hline
\hline
$\nth$ & $M_{\rm res}$ & $L_{\rm res}$ & $t_{\rm sim}$ & $dt_{\rm output}$ &  Fragment & $R_{\rm frag}$ & $t_{\rm form}$ & $t_{\rm merge}$ \\
($\cc$) & ($\msun$) & (au) & (yr) & (yr) & & (au) & (yr) & (yr) \\
\hline
$10^{15}$ (High) & 0.01 & 0.345 & 400 & 2 & H1 & 11 & 80 & 20 \\
& & & & & H2 & 24 & 158 & 18 \\
& & & & & H3 & 32 & 170 & 28 \\
& & & & & H4 & 19 & 240 & 20 \\
& & & & & H5 & 55 & 240 & 130 \\
& & & & & H6 & 48 & 300 & $>100$ \\
$10^{12}$ (Medium) & 0.1 & 7.43 & 7000 & 500 & M1 & 540 & 4000 & 1000 \\
& & & & & M2 & 315 & 4500 & 1500 \\
& & & & & M3 & 409 & 5500 & $>2000$ \\
& & & & & M4 & 700 & 6000 & $>1500$ \\
$10^{10}$ (Low) & 0.1 & 34.5 & 70000 & 2000 & L1 & $10^4$ & 0 & $>70000$ \\
\hline
\end{tabular}
\caption{
Column 1: Threshold number density (see Equ.~\ref{eq:ModelOpacity}),
Column 2: Resolution mass,
Column 3: Resolution length,
Column 4: Duration of simulation,
Column 5: Time span for data output,
Column 6: Number of fragments for each simulation,
Column 7: Separation from primary core when the fragment forms,
Column 8: Formation time after primary protostar formation, and
Column 9: Merger time of fragment to the primary.
Times for yet unmerged fragments are the survival times until the simulation ends.
}
\label{tab:summary}
\end{center}
\end{table*}

\subsubsection{Resolution limit of opaque core}

\cite{greif12} performed fully realistic simulations to determine the structure of the initial hydrostatic core (i.e. the protostar) in Pop~III star formation, employing a post-processing algorithm. Specifically, these authors
determine the spherically-averaged radius where the optical depth exceeds unity by considering the radial profile of the escape fraction around the protostar,
\begin{eqnarray}
\beta_{\rm esc}(R) = \frac{1}{N_{\rm ang}} \sum_j \frac{1 - \exp(-\tau_{j}(R))}{\tau_{j}(R)}~,
\label{eq:GreifEscape}
\end{eqnarray}
where $N_{\rm ang}$ is the number of angular bins, and the optical depth
\begin{eqnarray}
\tau_{j}(R) = \int^R_0 \rho_{j}(r) \kappa_{j}(r) dr~.
\label{eq:GreifOpacity}
\end{eqnarray}
Here, $\kappa_j$ denotes the Rosseland mean opacity.
The photospheric surface of the protostar, $R_{\rm ph}$, is identified as the location where the optical depth reaches unity, $\tau_j(R_{\rm ph}) = 1$, corresponding to $\beta_{\rm esc}(R_{\rm ph}) = 1 - \exp(-1) \simeq 0.63$
\citep[for further details, see sec.~2.7 in][]{greif12}.
The resulting protostellar radii range over $10 - 200~\rsun$, increasing with increasing protostellar mass.

Our opaque model leads to the formation of analogous hydrostatic cores.
Their surface can be defined by $\tau_{\rm art} = 1$, corresponding to a density contour with $\nh = \nth$.
Evidently, such artificial cores do not represent true protostars.
Even in our highest resolution simulation, the size of the artificially puffed-up hydrostatic core is $5~{\rm au} \sim 1000~\rsun$, which is much larger than the true core with $>10~\rsun$.\footnote{Note that the protostellar radius depends on the formation environment, in particular the mass accretion rate, e.g. $1~\rsun$ for the newborn hydrostatic core in \cite{yoshida08}.}
The former is comparable to the scale of the gravitationally unstable disc in \cite{greif12}.
Given our limited resolution, we cannot exclude the survival of protostars, once they have approached to within the size of the opaque core. 
Consequently, close binary systems may still be able to form \citep[e.g.][]{bate02}.
In addition, close encounters between protostars may result in their dynamical ejection from the cloud centre \citep[e.g.][]{greif11,smith11}.
The complicated dynamics inside the artificially puffed-up core may be important, but is beyond the scope of this paper.

\subsection{Local gravitational stability criterion} \label{sec:met_localQ}

To judge the occurrence of fragmentation in simulations, we introduce a local gravitational stability criterion for a rotating fluid.
A well-known stability criterion for differentially rotating discs was introduced by \cite{toomre64}, and is frequently used to understand simulation results \citep{boss97}.
Specifically, the Toomre criterion evaluates the parameter
\begin{eqnarray}
Q \equiv \frac{\cs \Omega}{\pi G \Sigma}~,
\label{eq:ToomreQ}
\end{eqnarray}
where $\cs$ is the sound speed, $\Omega$ the angular velocity, and $\Sigma$ the surface mass density. Stability then implies $Q>1$.
Here, we derive a similar criterion to evaluate instability for individual Lagrangean fluid particles. We thus do not pre-suppose axisymmetry, but can instead accommodate arbitrary geometries.

In general, a gaseous clump becomes gravitationally unstable when its self-gravity overcomes the opposing pressure and any support from shear motion.
The simplest case only considers the opposing pressure, leading to the classical Jeans criterion \citep{jeans1902}.
Let us now consider a region  with density $\rho$, sound speed $\cs$, and shear velocity $v_{\rm sh}$.
The gravitational contraction time is given by the free-fall time
\begin{eqnarray}
t_{\rm ff} = \sqrt{ \frac{3 \pi}{32 G \rho} } = 5.2 \times 10^7~{\rm yr}~\left( \frac{\nh}{\cc} \right)^{-1/2}~,
\label{eq:t_ff}
\end{eqnarray}
whereas the time for sound waves to cross the flattened (disc-like) system is
\begin{eqnarray}
t_{\rm sound} = \frac{H}{\cs}~,
\label{eq:t_sound}
\end{eqnarray}
with $H$ being the pressure scale height.
Therefore, the first condition for gravitational collapse is $t_{\rm ff} < t_{\rm sound}$.
Another requirement for instability is that the gravitational collapse proceeds faster than the timescale for shear motion to tear the system apart
\begin{eqnarray}
t_{\rm sh} = \frac{d}{v_{\rm sh}}~,
\label{eq:t_shear}
\end{eqnarray}
where $d$ is the overall scale of the collapsing region.
The second condition for instability is then $t_{\rm ff} < t_{\rm sh}$.
We can construct a combined criterion via $t_{\rm ff}^2 < t_{\rm sound} t_{\rm sh}$, resulting in
\begin{eqnarray}
\frac{\cs (v_{\rm sh} / d)}{\pi G (\rho H)} < \frac{32}{3 \pi^2} \sim 1~,
\label{eq:StableCritGeneral}
\end{eqnarray}
We evaluate the left-hand side of Equ.~(\ref{eq:StableCritGeneral}) for each SPH particle, allowing us to
examine the distribution of gravitational unstable regions in a three-dimensional setting.

In this study, fragmentation occurs in the rotating disc around the primary protostar.
We can thus optimize the stability criterion for our problem.
If we assume that the shear velocity is the rotational velocity around the disc centre, $v_{\rm shear} \to \vrot$, and the fragmentation scale is comparable to the radial distance from the centre, $d \to R$, the sound-crossing time can be written as $t_{\rm sound} \sim R/\vrot = \Omega^{-1}$.
By substituting this and $\rho H = \Sigma$ into Equ.~(\ref{eq:StableCritGeneral}), we recover Equ.~(\ref{eq:ToomreQ}).
Furthermore, by assuming the thin disc approximation, the pressure scale height is described as $H \sim R\cdot \cs/\vrot = \cs/\Omega$. Our local stability parameter (Equ.~\ref{eq:StableCritGeneral}) then becomes
\begin{eqnarray}
Q_{\rm local} \sim \frac{\Omega^2}{\pi G \rho}~.
\label{eq:StableCritGeneral-2}
\end{eqnarray}
In the next section, we show that the fragmenting regions in our simulations are well delineated by this criterion.

\begin{figure*}\begin{center}
\includegraphics[width=\textwidth]{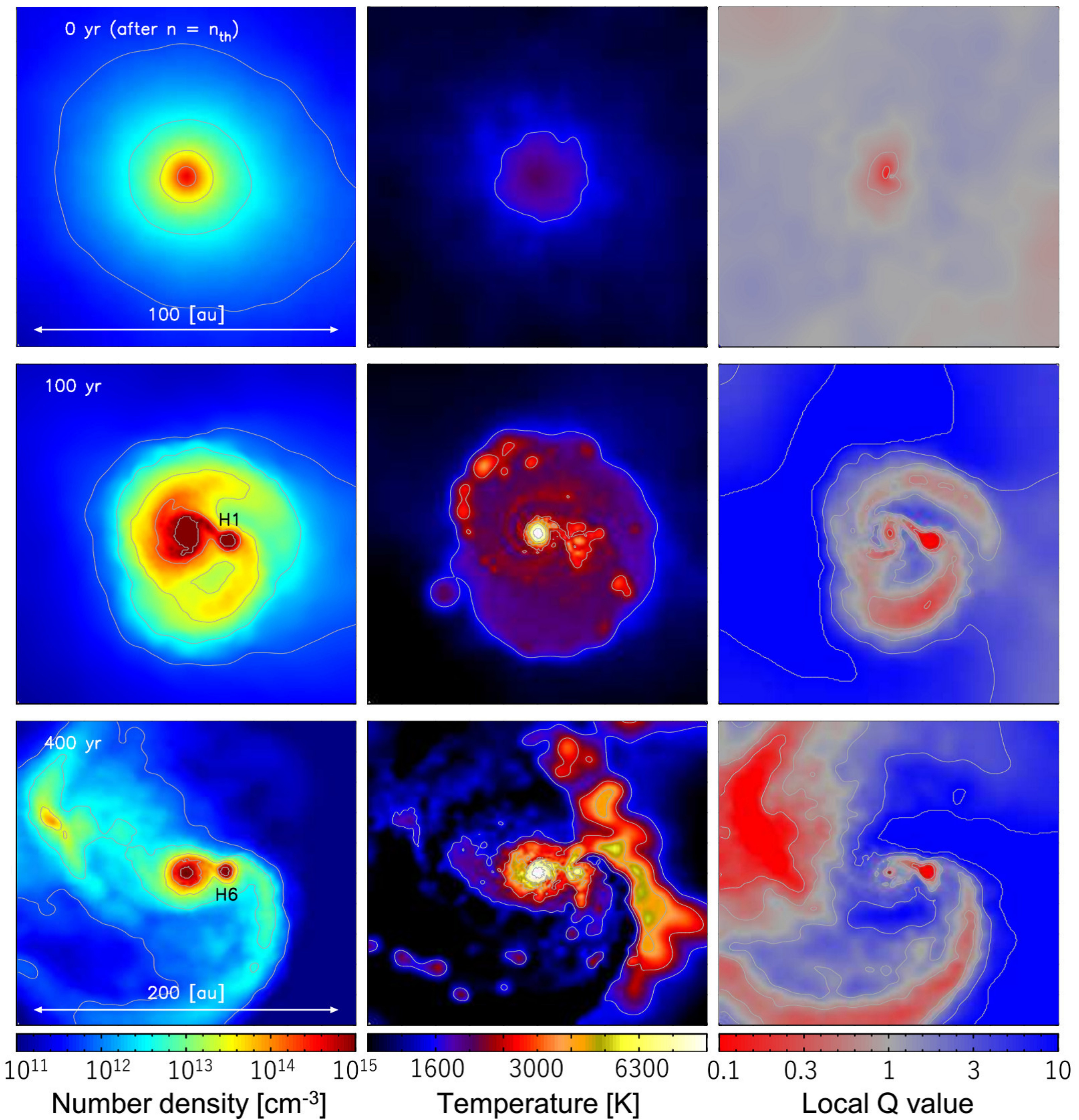}
\caption{
Cross-sectional view around the collapse centre of the primordial star-forming cloud in the high-resolution run with $\nth = 10^{15}~\cc$: gas number density (left panels), temperature (middle), and local Toomre stability criterion $Q$ (right) at 0, 100, and 400~yr after the primary protostar formation.
The box size is 100 (top and middle) and 200~au (bottom) on the side.
The labels (H1 and H6) indicate the corresponding fragments (Table~\ref{tab:summary}).
}
\label{fig:map1}
\end{center}\end{figure*}

\begin{figure*}\begin{center}
\includegraphics[width=\textwidth]{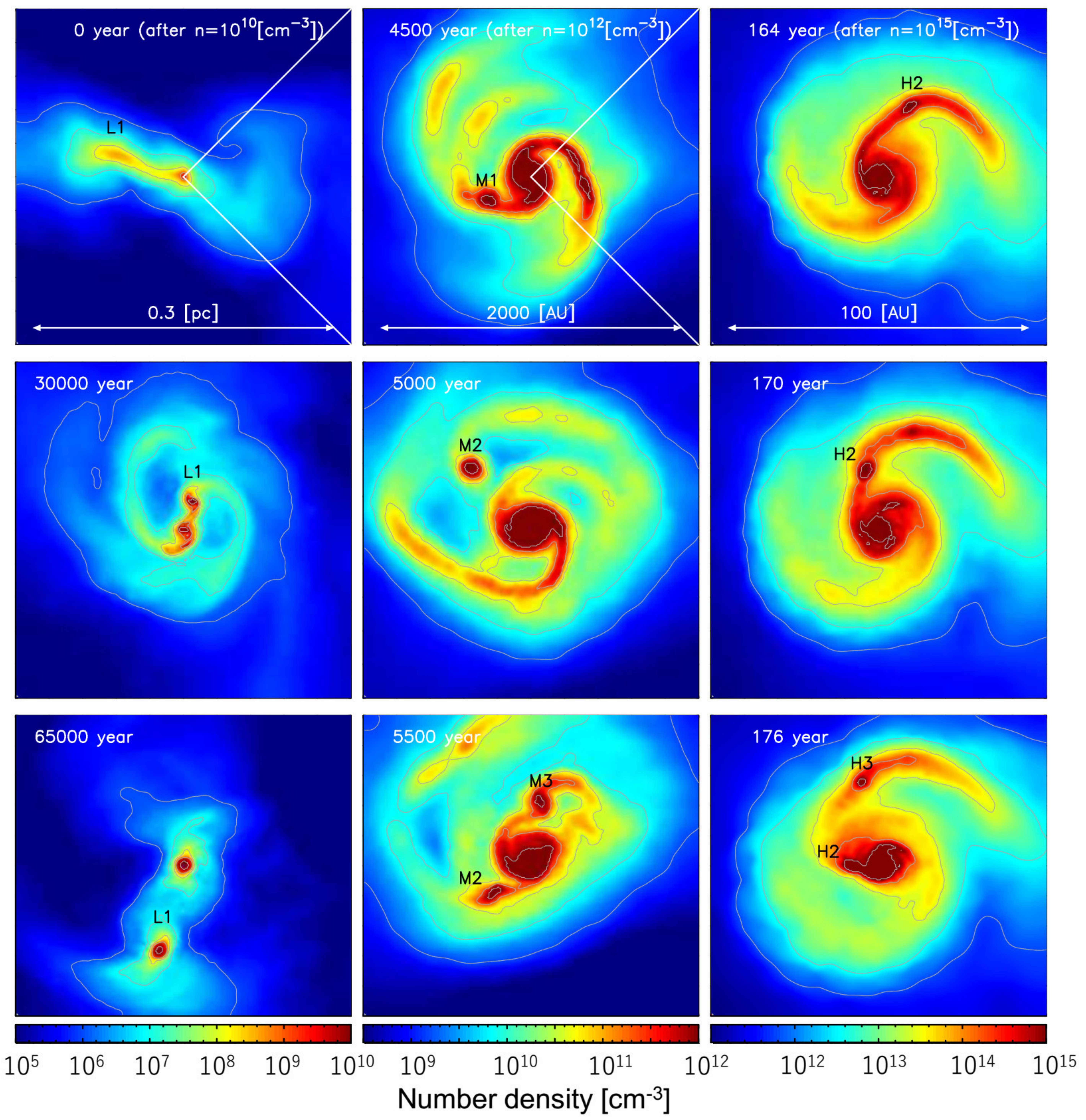}
\caption{
Cross-sectional view of the gas number density around the collapse centre of clouds.
Left, middle, and right panels shows results in the low-resolution run at 0, 30000, and 65000~yr,
medium-resolution run at 4500, 5000, and 5500~yr, and
high-resolution run at 164, 170, and 176~yr, respectively.
The box sizes are 0.3~pc, 2000, and 100~au, respectively.
Labels indicate the corresponding fragment (Table~\ref{tab:summary}).
}
\label{fig:map2}
\end{center}\end{figure*}

\section{Simulations} \label{sec:sim_res}

We now proceed to discuss our main simulation results. Specifically, we
perform three evolution calculations for a primordial star-forming cloud with different resolution scales, with threshold densities of $\nth = 10^{15}~\cc$ (High resolution run), $10^{12}~\cc$ (Medium), and $10^{10}~\cc$ (Low).
To control computational cost, simulation times are limited depending on the numerical resolution.
To explore the physics of fragment formation and survival, we continue the simulations until the formation and merger of a few fragments (for the High and Medium runs), or for about a contraction time of Pop~III stars (for the Low run).

\subsection{Initial collapse} \label{sec:sim_res_t0}

We first describe the initial cloud collapse phase, until primary protostellar core formation, marking the zero-point ($t = 0$~yr) for our subsequent discussion.
The newborn protostellar core is surrounded by rapidly accreting gas whose distribution is fitted by a power-law profile (Fig.~\ref{fig:Radi-Dens}).
In the top-row panels of Fig.~\ref{fig:map1}, we illustrate the physical state of the nearby material, by showing the gas number density (left panel), temperature (middle), and stability criterion (right).
At this time, the accretion disc exhibits a near-circular distribution of density and temperature.
The Toomre (local-$Q$) stability parameter indicates that only the cloud centre is gravitationally unstable ($Q < 1$), and that the extended disc is stable ($Q > 1$).
On larger scales, there is a second collapsing clump, located at $0.05~{\rm pc} = 10^4$~au from the primary protostar (see Fig.~\ref{fig:Radi-Dens}).
The top-left panel in Fig.~\ref{fig:map2} shows the density distribution around these two clumps. 
As can be seen, the clumps are formed inside a strongly elongated filament, which is in turn embedded in a Jeans unstable cloud with $R_{\rm J} = 0.8$~pc and $M_{\rm J} = 1500~\msun$.

\begin{figure}\begin{center}
\includegraphics[width=\columnwidth]{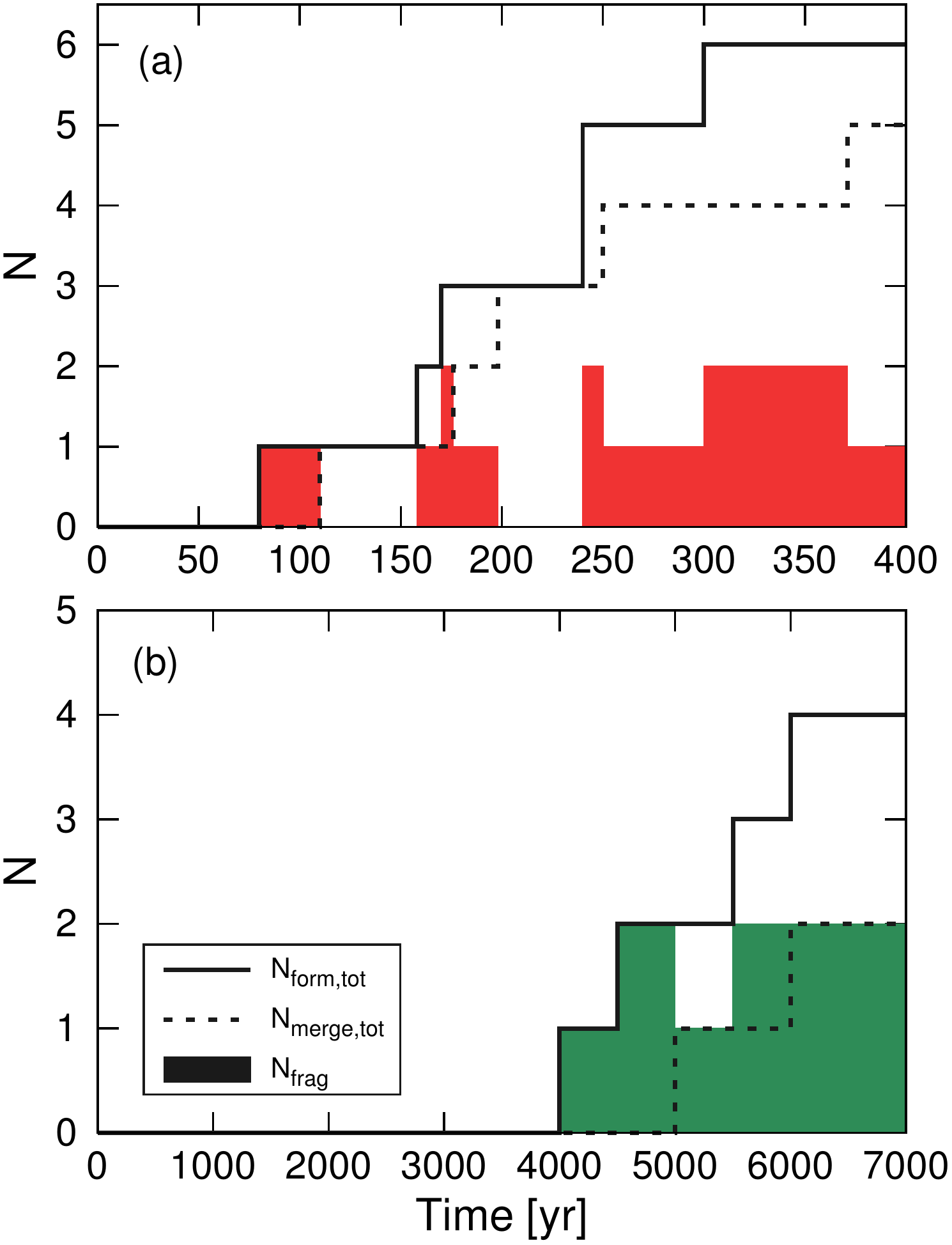}
\caption{
Time evolution of fragment number in High- (panel a) and Medium-resolution run (b).
The histograms show the distribution of surviving fragments.
The solid lines show the cumulative number of formed fragments, whereas the dashed ones depict that of merged fragments.
}
\label{fig:Time-Nfrag}
\end{center}\end{figure}

\begin{figure}\begin{center}
\includegraphics[width=\columnwidth]{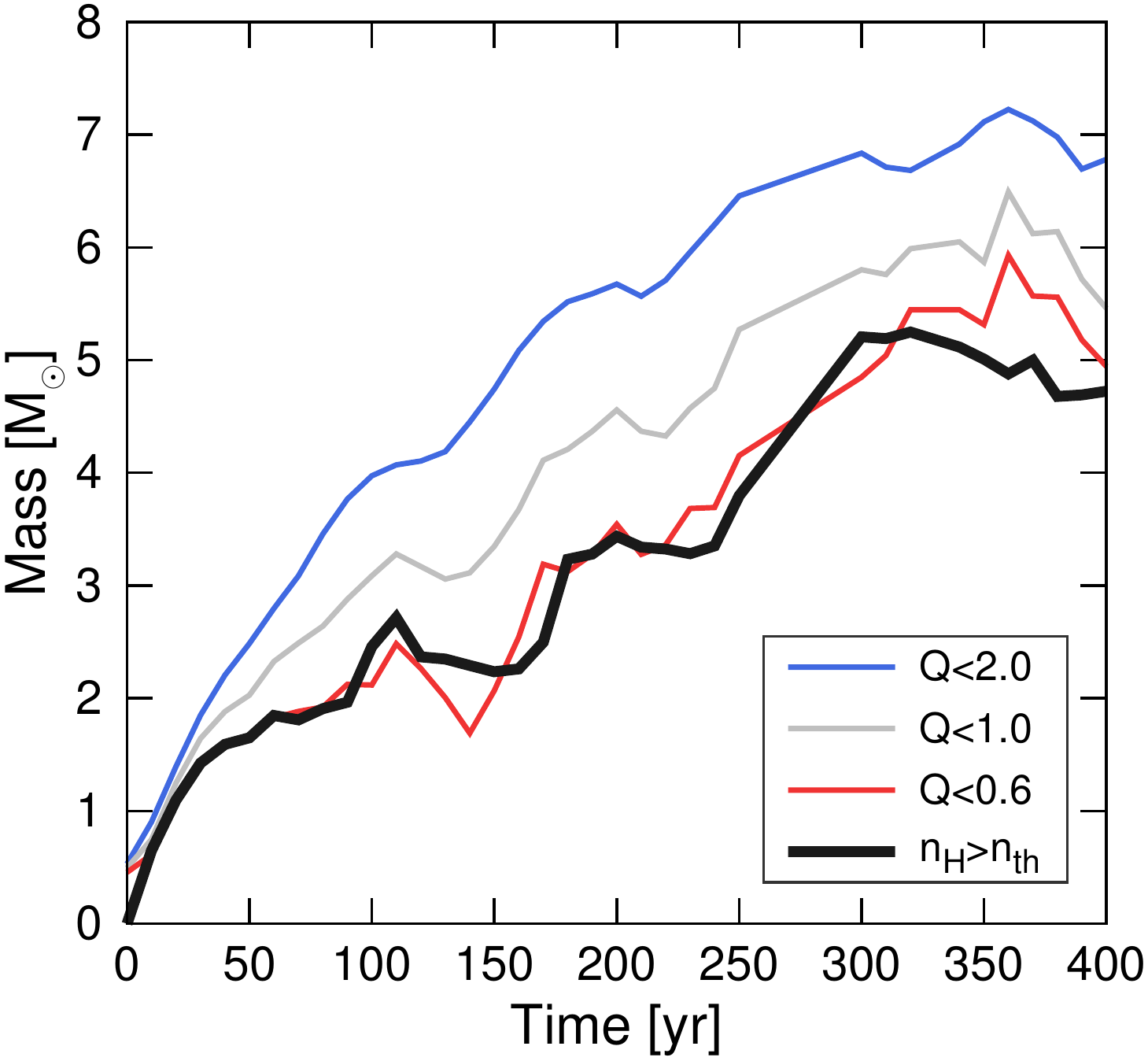}
\caption{
Total SPH particle mass, evaluated for different criteria (see below), in the high-resolution run as a function of time, measured after the gas density first reaches $\nth$.
The thick black line represents the mass of the hydrostatic core, where the gas number density exceeds $\nth$.
The thin coloured lines indicate the mass of SPH particles, where the local stability parameter is $Q < 2.0$ (blue), $1.0$ (grey), and $0.6$ (red), respectively.
}
\label{fig:Time-Mcore}
\end{center}\end{figure}

\subsection{High-resolution run} \label{sec:sim_res_t1}

For the high-resolution run, we set the threshold density to $\nth = 10^{15}~\cc$, corresponding to a resolution length of $0.345$~au and mass of $0.01~\msun$.
We terminate this simulation at $t = 400$~yr, a few times the free-fall timescale in the disc at $\sim 100$~au ($\sim 100$~yr).
In this run, six fragments form in the accretion disc, and five of them migrate to the centre, where they merge with the primary one.
In Fig.~\ref{fig:map1}, we show the time evolution of key cloud properties at $t =0$, $100$, and $400$~yr.
The initially smooth disc grows in mass via gas accretion, eventually forming spiral arms (left panels).
As a consequence of the accretion flow, the disc temperature increases
due to viscous heating (middle panels).
The massive disc is driven towards gravitational instability, with the
local stability parameter (Equ.~\ref{eq:StableCritGeneral-2}) delineating the boundary between the stable and unstable regions (right panels).

The first fragment (H1) forms in a spiral arm, located $11$~au from the primary protostar, at $t = 80$~yr.
It migrates inwards, while partaking in the general disc rotation, and finally merges with the central protostar $20$~yr after its appearance.
In Fig.~\ref{fig:map2} (in the panels of the right column), we show the subsequent formation and merger of fragments, specifically the second (H2) and third ones (H3), originating in the same spiral arm.
By the end of the simulation, three additional fragments have formed, all of which merge with the primary, except the final one.
The gravitationally unstable region becomes larger due to mass growth (right panels in Fig.~\ref{fig:map1}), so that the fragment scale, $R_{\rm frag}$, also increases (Table~\ref{tab:summary}).
In Fig.~\ref{fig:Time-Nfrag}(a), we summarize the resulting history of disc fragmentation. As can be seen, fragmentation occurs intermittently, but the number of surviving fragments does not increase due to migration-driven mergers.
We will more fully analyze this behaviour below (Section~\ref{sec:dis_angmom}).

In our numerical methodology, the surface of the puffed-up hydrostatic core is marked by the density contour at $\nth$.
We illustrate the mass growth history of the primary core in Fig.~\ref{fig:Time-Mcore}. 
During the first $400$~yr, the initial core grows to about $5~\msun$ via smooth gas accretion and through mergers with smaller fragments.
The rapid rise in mass at $t \sim 110$~yr is due to the merger with the H1 fragment.
After a brief stagnation phase, mass growth is reactivated when the H2 and H3 cores merge.
Towards the end of the run, the mass accretion rate onto the central blob is about $0.01~\msunyr$, which drives the rapid growth of the central protostar inside the blob.
We also plot the total masses of particles whose local stability parameter is less than a certain level.
Here, the mass of the central core is comparable to that of SPH particles with $Q<0.6$, and less than of those with $Q < 1$, where the disc fluid becomes gravitationally unstable.
This is further illustrated in Fig.~\ref{fig:map2}, where the spiral arm structure corresponds to the contour with $Q = 1$, and fragmentation occurs in more unstable regions inside the spiral arm where $Q < 1$ \citep[see also][]{takahashi16}.

\subsection{Medium-resolution run} \label{sec:sim_res_t2}

The medium-resolution run with $\nth = 10^{12}~\cc$ can only resolve scales of $7.43$~au and $0.1~\msun$, but continue for a longer time than the high-resolution one.
More specifically, this run explores the fragmentation history of the larger-scale disc, out to a radius of $1000$~au.
The overall phenomenology is similar to the more resolved run.
In the middle column of Fig.~\ref{fig:map2}, the fragmentation inside the main spiral arms is evident.
After $7000$~yr, four fragments have formed at a similar initial distance of a few $100$~au, with the first two migrating into the primary core within $1000 - 1500$~yr.

Fig.~\ref{fig:Time-Nfrag}(b) shows the formation and merger histories of fragments in this run, to be compared with the high-resolution results above.
Here, the onset of fragmentation is shifted to later times, as a consequence of the increase in length scales that can be resolved. Our series of runs with varying resolution thus explores inside-out disc fragmentation, as it unfolds in time.
The timescale for the onset of fragmentation in each case is a few times larger than the free-fall time at the respective fragmentation scale, $t_{\rm ff}(R_{\rm frag})$, in agreement with what was found in previous studies \citep{greif12,stacy16}.

In the medium-resolution run, the secondary clump (corresponding to L1) is also able to collapse, with multiple fragments forming inside it. 
Again, the fragmentation is triggered within spiral arms, and three fragments survive at the end of the simulation. 
We thus have a hierarchically structured situation, similar to what is seen in present-day star formation.

\subsection{Low-resolution run} \label{sec:sim_res_t3}

Finally, we discuss the third run with $\nth = 10^{10}~\cc$ to investigate the long-term fate of the pre-collapsed distant clump, located $10^4$~au from the primary protostar.
The initial orbital timescale of the distant clump is $(R^3/G M)^{1/2} \sim 10^4$~yr, where $R$ is the clump radius and $M$ its mass.
We continue this run for $7 \times 10^4$~yr, which is longer than the orbital time, and corresponds to the ZAMS contraction time (Section~\ref{sec:the_bin_survival}).

\begin{figure}\begin{center}
\includegraphics[width=\columnwidth]{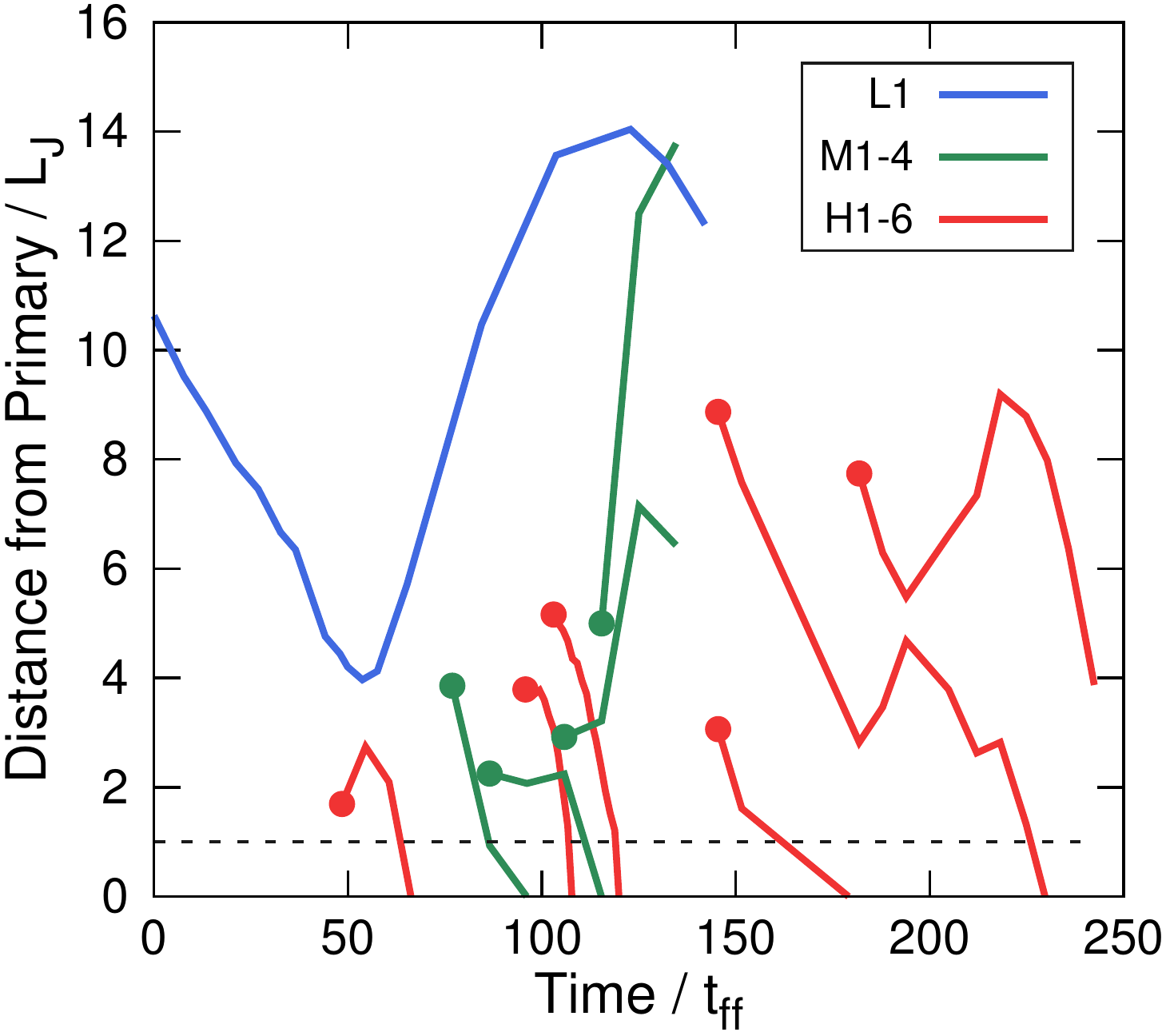}
\caption{
Distance of fragments from the disc centre as a function of time after the central density first reaches $\nth$.
Time and distance are normalized by characteristic scales, evaluated for the opaque core, with temperatures of $900$, $1000$, and $2000$~K at $\nh = 10^{10}$, $10^{12}$, and $10^{15}~\cc$, respectively. 
This results in free-fall times of $t_{\rm ff} = 520$, $52$, and $1.7$~yr, and Jeans lengths of $L_{\rm J} = 1300$, $140$, and $6.2$~au, respectively.
The coloured lines represent the three simulation cases, with solid circles indicating when and where fragments form.
The horizontal dashed line indicates a Jeans length, in other words, the approximate radius of the primary opaque core (`protostar').
}
\label{fig:Time-Dist}
\end{center}\end{figure}

In Fig.~\ref{fig:Time-Dist}, we show the distance evolution of these two clumps (blue line).
As can be seen, the secondary clump cannot directly migrate to the primary one. 
To assess its further evolution, we calculate the timescale of angular momentum redistribution, evaluated at the time of primary protostar formation, as a guide,
\begin{eqnarray}
t_{\rm ang} = \frac{|{\bf l}|^2}{{\bf l} \cdot {\bf \tau}} \sim 5 \times 10^7~{\rm yr}~.
\label{eq:t_ang}
\end{eqnarray}
Here ${\bf l} = \sum {\bf r}_i \times (m_i {\bf v}_i)$ is the angular momentum, ${\bf \tau} = \sum {\bf r}_i \times (m_i {\bf a}_i)$ the torque, and $m_i$, ${\bf v}_i$, and ${\bf a}_i$ are the mass, velocity, and acceleration of an individual SPH particle.
This timescale is to be compared to the assembly time for the central protostar.
Considering the protostellar radiation feedback \citep{mckee08}, this growth history was examined in a previous study by performing a two-dimensional axysymmetric radiation hydrodynamic simulation \citep{hirano14}.
The final stellar mass was $380~\msun$, reached $0.14$~Myr after initial protostellar core formation. Consequently, when the primary protostar
reaches the ZAMS, and when its radiation feedback is activated (at 0.14~Myr), the secondary clump is predicted to maintain a rather wide orbital distance.

\section{Methodological Limitations} \label{sec:caveat}
Before further interpreting our simulation results, we briefly address some key limitations inherent to our numerical methodology.
In this study, we introduce an artificial hydrostatic core to arrest the gravitational collapse at a pre-determined resolution scale, thus making it possible to follow the long-term evolution of the system.
By using this approach, instead of the widely used sink particle method, our calculations naturally treat the viscous diffusion occuring during close encounters of gas clumps. Such viscous friction may lead to the merger of the clumps.\footnote{
Newer sink algorithms \citep{hubber13} adopt a hybrid strategy, combining the sink particle and hydrostatic core methodologies, by preserving a `working fluid' inside the sink particle that can react to the surrounding disc material and provide a reservoir into which collisional energy can be dissipated.
The idea is that the sink particle should only incorporate material at the very centre of the collapsing fragment, leaving the surrounding gas to interact with the rest of the fluid in the computational volume.}
However, the artificial core model cannot properly resolve sub-grid effects, giving rise to a number of possible problems, as follows:
\begin{itemize}
\item Our methodology tends to smooth out the sharpness of any spiral arm structures.
This implies that the disc is artifically rendered denser overall -- whereas when spiral arms are sharp, it creates voids of lower density material in the disc, in turn lowering the efficiency of viscous transport.
Also, sharper spiral arms are less able to channel their fragments towards the centre.
As an example, the density contours in Fig.~\ref{fig:map2} illustrate this suppression of the density contrast within the disc/spiral arm structure.
\item Furthermore, the artificial enhancement of the object size softens close dynamical encounters.
Such close encounters led to the dynamical ejection of a significant
fraction of the fragments in the
simulations of \cite{greif12}, where the true protostellar scale was resolved. 
\end{itemize}
Consequently, our adopted methodology may thus
strongly increase the impact of viscous transport, while at the same time dramatically reducing the incidence of dynamical ejections.
In effect,
our calculations with the stiffened equation of state investigate the opposite simplification to that made in the sink particle method, and reality is bracketed
by these two approaches.

\section{Implications and Analysis} \label{sec:dis}

The purpose of this study is to assess the multiplicity of Pop~III stellar systems, focusing on the survival conditions of migrating protostellar fragments (Section~\ref{sec:the_bin_survival}).
The three simulations presented here illustrate fragmentation at different scales and their respective dynamical and viscous evolution. 
The latter results either in coalescing of fragments or their long-term survival.
In this section, we discuss the physics of disc fragmentation and viscous transport in greater detail, thus providing an improved understanding of whether Pop~III binaries and multiples can form and survive.

\subsection{Angular momentum redistribution}\label{sec:dis_angmom}

\begin{figure}\begin{center}
\includegraphics[width=\columnwidth]{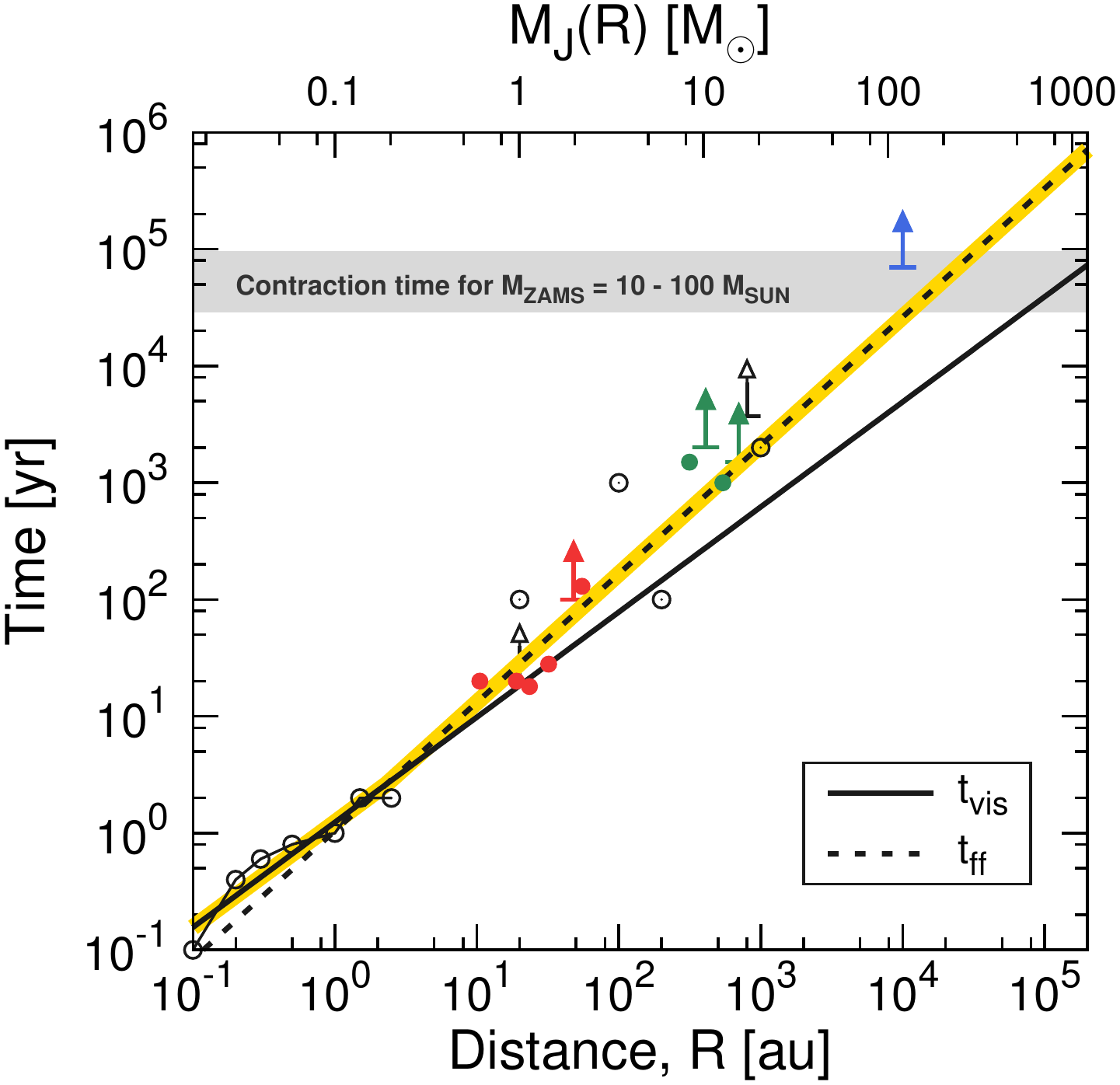}
\caption{
Fragment migration time vs. distance $R$ from the primary protostellar core.
Filled circles and arrows indicate the current simulation results (Table~\ref{tab:summary}), whereas unfilled circles \citep{greif11,greif12,stacy16,hosokawa16,stacy13b} and arrows \citep{clark11b,turk09} represent previous studies, in order of lower- to higher-$R$, respectively.
The circles show exact migration times, whereas arrows mark lower limits because fragments have not yet merged.
The solid and dashed lines are the viscous (Equ.~\ref{eq:t_merger-2}) and free-fall timescales (Equ.~\ref{eq:t_ff-2}), and the thick yellow line shows the maximum of them.
The grey band represents the range of contraction times for the evolution of protostars with $M_{\rm ZAMS} = 10 - 100~\msun$, in other words, the time required for strong radiative feedback to be launched.
The top scale represents the Jeans mass at the corresponding scale (Equ.~\ref{eq:Mieans-Radi}), which is very similar to the enclosed mass (Equ.~\ref{eq:Menc-Rfrag}).
}
\label{fig:Rfrag-Time}
\end{center}\end{figure}

In Fig.~\ref{fig:Rfrag-Time}, we summarize how migration times in the protostellar disc depend on initial fragmentation scale, with distance measured from the primary protostar. 
To cover a large range of scales, we complement our results with those from previous 3D (radiation-)hydrodynamic simulations \citep{clark11b,greif11,greif12,stacy13b,stacy16,hosokawa16}.
The numerical data points suggest a definite correlation that can be understood as follows.
\cite{greif12} argues that the resulting migration to the cloud centre (open circles connected with a line at left-bottom in Fig.~\ref{fig:Rfrag-Time}) typically occurs in a free-fall time, because the angular momentum redistribution via gravitational torques is so efficient.
Assuming a power-law density distribution with parameters $A$ and $b$, such that $\nh(R) = A/\cc~(R/{\rm au})^{-b}$, the free-fall time (Equ.~\ref{eq:t_ff}) can be rewritten as a function of distance $R$,
\begin{eqnarray}
t_{\rm ff} &=& 5.2 \times 10^7~{\rm yr} \cdot A^{-1/2} R^{b/2}~,\\
&\sim& 1.05~{\rm yr}~(R/{\rm au})^{1.1}~.
\label{eq:t_ff-2}
\end{eqnarray}
Here, we have substituted a density distribution of $\nh \sim 2.5 \times 10^{15}~\cc~(R/{\rm au})^{-2.2}$.
The dotted line in Fig.~\ref{fig:Rfrag-Time} successfully reproduces data in \cite{greif12}, but also other data on larger scales.

To elucidate the origin of this clear relationship, we evaluate the viscous timescale in the disc, assuming typical properties obtained from primordial star-formation simulations.
Within the thin disc approximation, the $\alpha$-disc model \citep{shakura73} describes the viscous timescale as a function of the distance from the disc centre as
\begin{eqnarray}
M_{\rm disc} &\sim& \pi R^2 \Sigma~,\\
\frac{dM_{\rm disc}}{dt} &\sim& 3 \pi \nu \Sigma~,\\
t_{\rm vis} &=& \frac{M_{\rm disc}}{dM_{\rm disc}/dt} \sim \frac{R^2}{3 \nu} \sim \frac{R^2 \Omega}{3 \alpha \cs^2}~,
\label{eq:t_merger-1}
\end{eqnarray}
where $\nu$ ($= \alpha \cs^2 / \Omega$) is the viscous coefficient.
The radial dependence of the enclosed gas mass, characterized by a power-law density distribution, is described as
\begin{eqnarray}
M_{\rm enc}(R) &=& \int_0^{R} 4 \pi r^2 \mh \nh(r) dr~,\\
&=& 3.51 \times 10^{-17}~\msun \cdot \frac{A}{3-b} R^{3-b}~,\\
&\sim& 0.11~\msun~(R/{\rm au})^{0.8}~.
\label{eq:Menc-Rfrag}
\end{eqnarray}
The accretion disc rotates nearly Keplerian with $f_{\rm Kep} = \vrot / \vkep = (R^3 \Omega^2 /G M)^{1/2} \sim 0.5$, resulting in
\begin{eqnarray}
\Omega(R) &=& \frac{1}{2} \left( \frac{G M}{R^3} \right)^{1/2}~,\\
&=& 1.86 \times 10^{-8}~{\rm yr^{-1}} \cdot \left( \frac{A}{3-b} \right)^{1/2} R^{-b/2}~,\\
&\sim& 1.03~{\rm yr^{-1}}~(R/{\rm au})^{-1.1}~.
\label{eq:Omega}
\end{eqnarray}
By substituting these expressions, the viscous timescale becomes
\begin{eqnarray}
t_{\rm vis} &\sim& 3.76~{\rm yr}~(R/{\rm au} \cdot M_{\rm enc}/\msun)^{1/2} \notag \\
&& \cdot \left( \frac{\alpha}{1.0} \right)^{-1} \left( \frac{\cs}{2.5~\kms} \right)^{-2} \, ,\\
&=& 2.23 \times 10^{-8}~{\rm yr} \cdot \left( \frac{A}{3-b} \right)^{1/2} R^{(4-b)/2} \notag \\
&& \cdot \left( \frac{\alpha}{1.0} \right)^{-1} \left( \frac{\cs}{2.5~\kms} \right)^{-2} \, ,\\
&\sim& 1.24~{\rm yr}~(R/{\rm au})^{0.9} \notag \\
&& \cdot \left( \frac{\alpha}{1.0} \right)^{-1} \left( \frac{\cs}{2.5~\kms} \right)^{-2}~.
\label{eq:t_merger-2}
\end{eqnarray}
This timescale depends on the physical properties of the disc, specifically its density and temperature distribution.

For gravitational torques, the Shakura-Sunyaev parameter has a typical value of $\alpha\simeq 1.0$, and the sound speed is almost constant after the loitering phase, about $2 - 3~\kms$ for this cloud.
For these conditions, the resulting expression indicates that $t_{\rm vis} \le t_{\rm ff}$ in the primordial protostellar disc, with a radial extent of $R < 10^4$~au.
The timescale for the fragment merger process is restricted to the longer of the two (yellow line in Fig.~\ref{fig:Rfrag-Time}).
So, in the primordial disc, fragments lose angular momentum and migrate to the centre on approximately the dynamical time.\footnote{The comparison between the two timescales depends on the disc properties. We show the most distinct case for a Pop~I disc, where the viscous time becomes longer than the free-fall time, in Section~\ref{sec:dis_PopI} (see Fig.~\ref{fig:Rfrag-Time-PopI}).}
To examine whether fragments, formed at a certain scale, will likely survive or merge, it is only necessary to continue the simulation over the timescale that corresponds to the respective fragmentation scale.
This timescale consideration can be affected by the evolving density distribution in the disc, given the strong dependence on the $A$ parameter.
The growth of the disc mass results in an increasing $A$, which in turn implies an increase in $t_{\rm vis}$, whereas $t_{\rm ff}$ decreases.

\subsection{Radiative feedback}

The migration of fragments via angular momentum redistribution by gravitational torques can continue until the gaseous component is blown away by the stellar radiative feedback.
The radiation feedback becomes effective when the star reaches the ZAMS, after a contraction time from protostellar core formation of $t_{\rm cont} = 3 \times 10^4 - 10^5$~yr for $10 - 100~\msun$ Pop~III stars (Section~\ref{sec:the_bin_survival}).
For a fragment to move close to the primary star within this time, we have the condition $t_{\rm ff} \le t_{\rm cont}$, which restricts the fragmentation scale to about $R \le (1 - 3) \times 10^4$~au (intersection between the yellow line and grey band in Fig.~\ref{fig:Rfrag-Time}).
This is smaller than the scale of the initial Jeans-unstable parent cloud of $\sim 6 \times 10^5$~au (Equ.~\ref{eq:Ljeans}).
To estimate the mass scale for fragmentation at a given radial distance, we substitute our power-law density profile into the equation for the Jeans mass (Equ.~\ref{eq:Mjeans-Dens}),
\begin{eqnarray}
M_{\rm J} &\approx& 0.15 (A/10^{15}~\cc)^{-0.35} (R/{\rm au})^{-0.35 b} \, ,\\
&\sim& 0.11~\msun~(R/{\rm au})^{0.77}~,
\label{eq:Mieans-Radi}
\end{eqnarray}
which is very similar to the enclosed mass distribution (Equ.~\ref{eq:Menc-Rfrag})

Consequently, the most massive gravitationally unstable fragment which can approach to the primary core until radiative feedback photo-evaporates the gaseous disc medium has $M_{\rm J} \sim 100 - 300~\msun$.
Inside such a massive fragment, the formation of massive black holes with $\sim 30~\msun$ is plausible.
If the viscous dissipation could lessen the distance between primary and fragment clumps to less than $0.1 - 1$~au, close enough for possible BH remnants to merge within the Hubble time, the final BH merger could produce an observable GW signal (Section~\ref{sec:the_bin_ms}).

In addition to the radiative effect from the primary star, secondary fragments might contribute to photo-evaporating the accretion disc if they contract to the main sequence before migrating to the primary protostar \citep[fig.~3 in][]{inayoshi14}.
However, assuming that fragments collapse on the local free-fall time, the timescale for migration, given by the viscous time, is similar, $t_{\rm ff} \sim t_{\rm vis}$, as we have shown above.
Any secondary protostars thus cannot ``outrun'' the primary, in terms of triggering radiative feedback.
This is the same conclusion reached in previous studies, based on analytical models of the accretion disc around a protostar \citep{inayoshi14,latif15}; {\it clumps will migrate inward before they reach the main sequence and produce UV feedback} \citep{latif15}.

\subsection{Comparison with Population~I}\label{sec:dis_PopI}

In the discussion above, we have obtained the characteristic scale where the migration timescale, $t_{\rm mig} = \max \{{t_{\rm ff}, t_{\rm vis}}\}$, becomes equal to the protostellar contraction timescale, after which radiation feedback limits protostellar mass growth, $t_{\rm cont}$.
As we have seen, for the Pop~III case this critical scale is of the same order as the protostellar disc size, $\sim 10^4$~au.
Fragments formed anywhere in the disc, therefore, can lose angular momentum via viscous torques and migrate to the centre, before the gaseous medium is blown away via radiation feedback.

\begin{figure}\begin{center}
\includegraphics[width=\columnwidth]{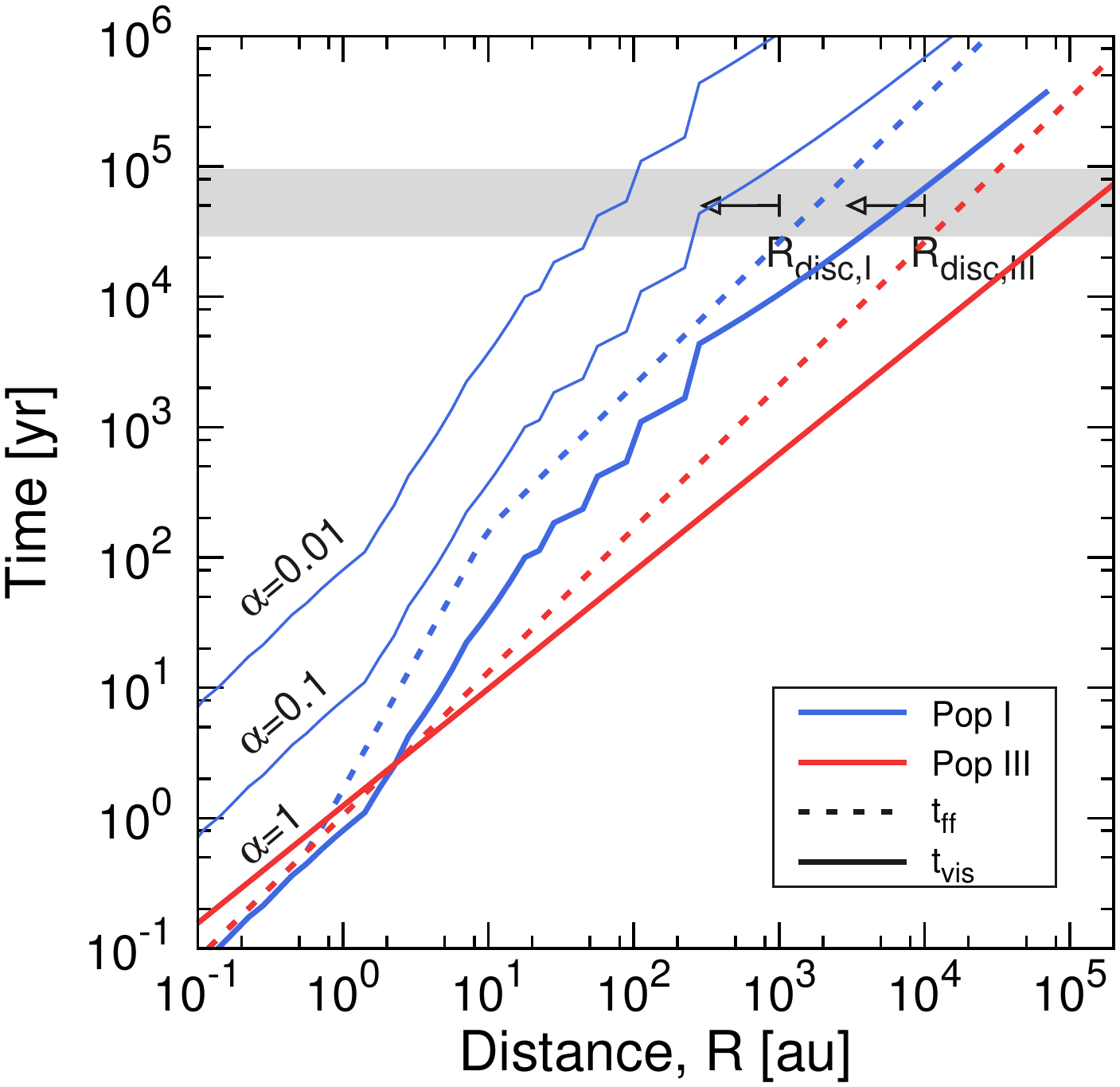}
\caption{
Timescales for Pop~I case as a function of distance from the disc centre.
The solid and dashed lines show the viscous timescales and free-fall timescales for Pop~I (blue; with $\alpha = 1$, $0.1$, and $0.01$) and Pop~III (red) cases, respectively.
The grey region represents the range of evolutionary contraction times for protostars with $M_{\rm ZAMS} = 10 - 100~\msun$.
The arrows mark the typical disc radius for the Pop~I and III cases, $10^3$ and $10^4$~au, respectively.
}
\label{fig:Rfrag-Time-PopI}
\end{center}\end{figure}

How does this situation compare to the present-day, Population~I (Pop~I), case?
Applying the same methodology as above to solar-metallicity discs \citep[see][for a recent review]{kratter16}, the relevant timescales (Eqs.~19 and 36) are controlled by the density $\nh$, the sound speed $\cs$ (i.e. temperature), and the viscous parameter $\alpha$.
We adopt the hydrodynamical calculation results of gravitationally collapsing clouds with solar metallicity in \cite{omukai10}.
The fitted functions of density and temperature distributions are summarized in Appendix~\ref{app:PopIdisc} and Fig.~\ref{figA:DiscModel}.
The sound speed depends on gas temperature as
\begin{eqnarray}
\cs &=& \sqrt{\gamma k_{\rm B} T / m_{\rm H} \mu} \, ,\\
&\simeq& 2.5~{\rm km~sec^{-1}} \cdot \left( \frac{T}{1000~{\rm K}} \frac{\gamma}{5/3} \frac{2.4}{\mu} \right)^{1/2}~,
\label{eq:SoundSpeed}
\end{eqnarray}
where $\gamma$ is the ratio of specific heats, and $\mu$ the mean molecular weight.
Finally, considering a gravitationally unstable disc and its fragmentation, we again adopt $\alpha \simeq 1$ as fiducial viscous parameter, the same as for the Pop~III case above.
By substituting these values into Equ.~(\ref{eq:t_ff}) and (36), we obtain the free-fall and viscous timescales for the Pop~I case.

In Fig.~\ref{fig:Rfrag-Time-PopI}, we plot the resulting viscous and free-fall timescales. 
With the fiducial viscous parameter ($\alpha = 1$), $t_{\rm vis} \le t_{\rm ff}$ on all scales for the Pop~I case, similar to the Pop~III one.
The increase in both timescales for the Pop~I case, compared to Pop~III, is due to the lower density and temperature. Specifically, the two timescales scale as $t_{\rm ff}(R) \propto \nh(R)^{-1/2}$ and $t_{\rm vis}(R) \propto \nh(R)^{1/2} R^2 T(R)^{-1}$, respectively.
The effective migration scale, where $t_{\rm mig} = \max(t_{\rm ff}, t_{\rm vis})$ falls below the protostellar contraction time, is of order the typical Pop~I disc size, $100 - 1000$~au, such that $R_{\rm mig}/R_{\rm disc} \sim 1$, similar to the situation in a Pop~III disc.
Consequently, all fragments formed in the disc can migrate, with typically Jeans masses of $0.1$ and $0.6~\msun$ at $R = 100$ and $1000$~au.
If the viscous efficiency were diminished to $\alpha = 0.01 - 0.1$ for some reason, the viscous timescale would be longer than the free-fall one.
The migration scale would then become less than the disc size, and fragments originating beyond this scale would survive without migration.

\begin{figure}\begin{center}
\includegraphics[width=\columnwidth]{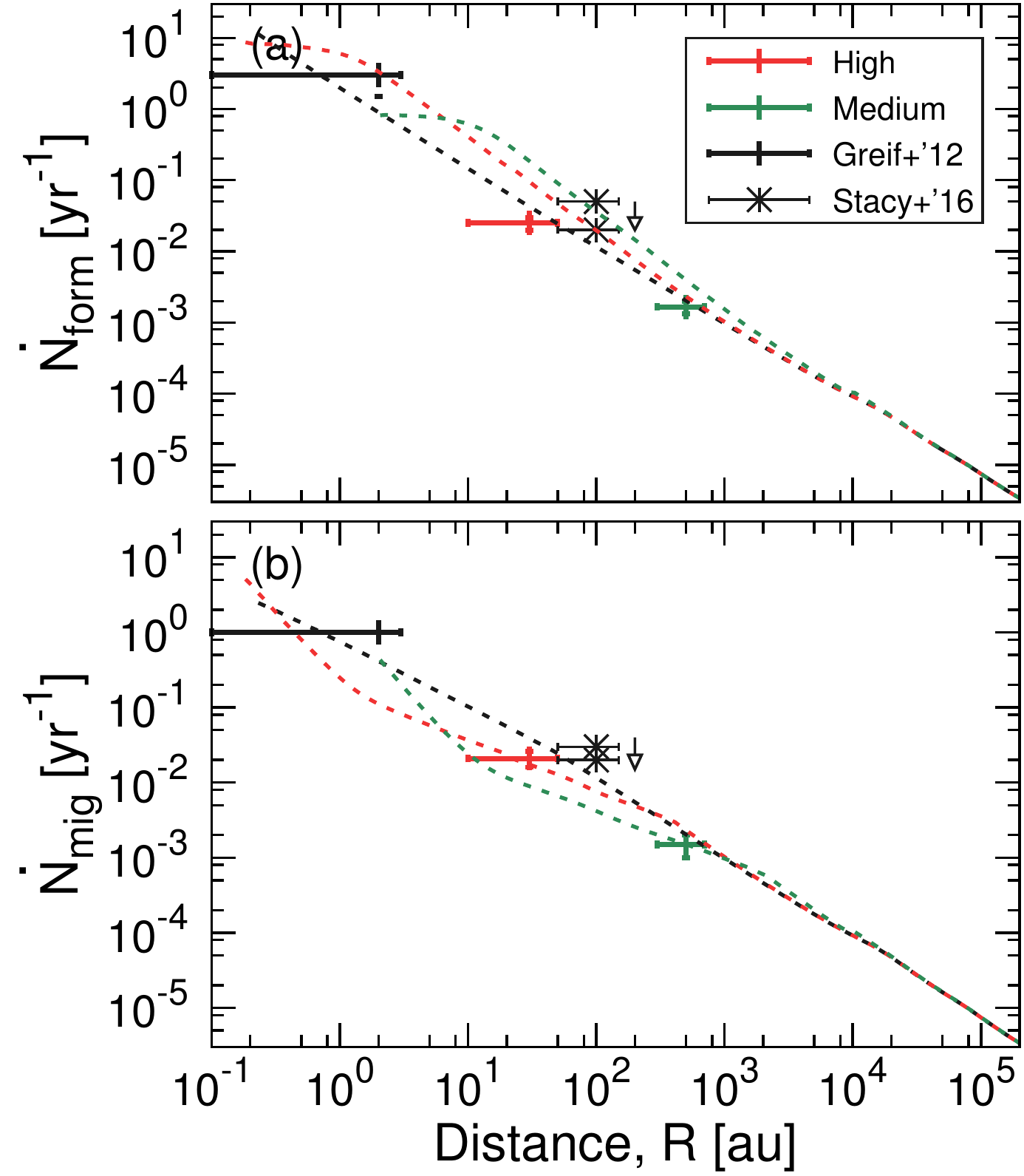}
\caption{
Shaping the mass spectrum of Pop~III stars.
{\it Panel a:} Formation rate of fragments as a function of distance from the disc centre \citep[for simulations in this study and in][]{greif12,stacy16}.
{\it Panel b:} Corresponding migration rate.
The dashed lines show analytical rate estimates, calculated from the local gas properties, at $t = 0$ (black), $400$ (red), and $7000$~yr (green). 
The star symbols, connected with arrows, illustrate the decline in the formation rate after $2000$~yr \citep[seen in][]{stacy16} which in turn reflects the depletion of the available disc gas.
}
\label{fig:Rfrag-dNdt}
\end{center}\end{figure}

\subsection{Building-up the IMF} \label{sec:dis_IMF}

What are the implications for the assembly of the Pop~III IMF?
We will follow the approach in \cite{dopcke13}, who compare two timescales, that for fragmentation, $t_{\rm frag} \equiv M_{\rm J}/\dot{M}$, and that for accretion, $t_{\rm acc} \equiv M_{\rm enc}/\dot{M}$, where $\dot{M}$ is the accretion (or infall) rate.
If the former is larger than the latter, the newly formed fragments are quickly accreted onto the primary protostar, so that more massive star formation is promoted.
In the opposite case, fragment formation occurs more rapidly than the merger process, such that lower-mass star formation dominates.
They conclude that the transition of $t_{\rm frag}/t_{\rm acc} > 1$ to $< 1$ indicates the transformation of the stellar mass distribution from high-mass to low-mass dominated, and their analysis can reproduce the behaviour seen in the simulations.

In Fig.~\ref{fig:Rfrag-dNdt}, we summarize the formation and migration rates of fragments at different fragmentation scales, obtained for the high- and medium-resolution runs in this study, as well as the \cite{greif12} and \cite{stacy16} simulations. 
We also illustrate the time evolution of $\dot{N}_{\rm form} = t_{\rm ff}^{-1}$ and $\dot{N}_{\rm mig} = t_{\rm mig}^{-1} = 1/\max\{t_{\rm ff}, t_{\rm vis}\}$, employing the local conditions in our simulated discs, at three times, $t=0$, $400$, and $7000$~yr after primary protostar formation.
This analytical estimate is in good agreement with the actual behaviour seen in the simulations. 
Considering the evolution of the analytical profiles, the formation rate seems to increase with time, whereas the migration rate decreases, which would imply an increasing number of surviving fragments. 
However, there is another important effect to properly gauge the time evolution of the fragment formation rate, related to the depletion of the available gas reservoir.
This effect is responsible for the slight mismatch between the simulation results and the locally evaluated rates (see Fig.~\ref{fig:Rfrag-dNdt}), and for the decrease in the formation rate, as seen in the \cite{stacy16} study.
With these rates in hand, we are now in a position to evaluate the resulting build-up of the Pop~III IMF, which will reflect their relative importance as a function of time.

\begin{figure}\begin{center}
\includegraphics[width=\columnwidth]{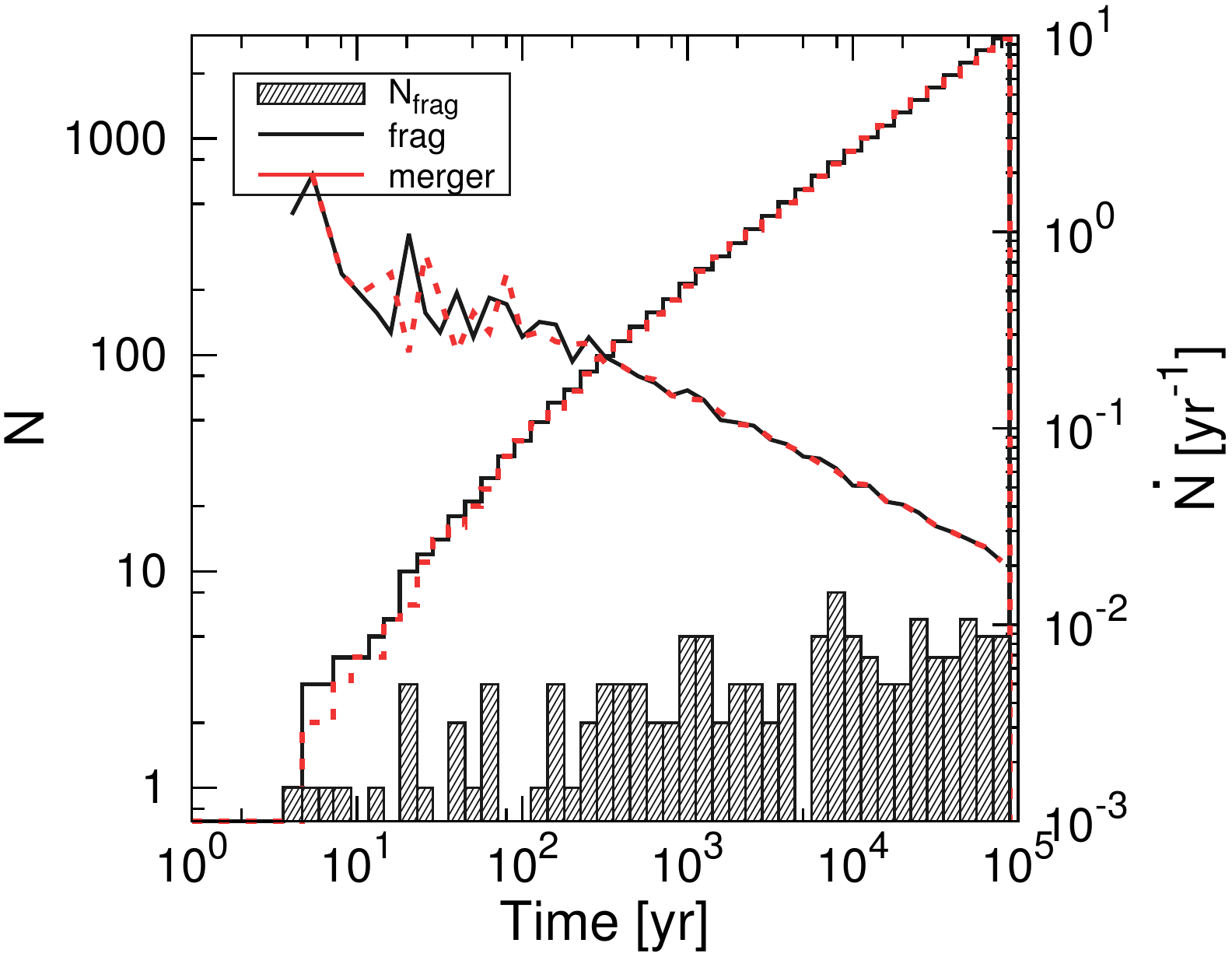}
\caption{
Time evolution of fragment number in the primordial protostellar disc, according to analytical model.
The spectrum shows the number of surviving fragments, as a function of time after primary protostar formation.
The steps represent the cumulative number of formed (solid black) and merged (dashed red) fragments.
In addition, the lines show the corresponding rates of fragment formation (solid black) and mergers (dashed red), both of which decline with time.
}
\label{fig:Time-N+dNdt}
\end{center}\end{figure}

\begin{figure}\begin{center}
\includegraphics[width=0.45\textwidth]{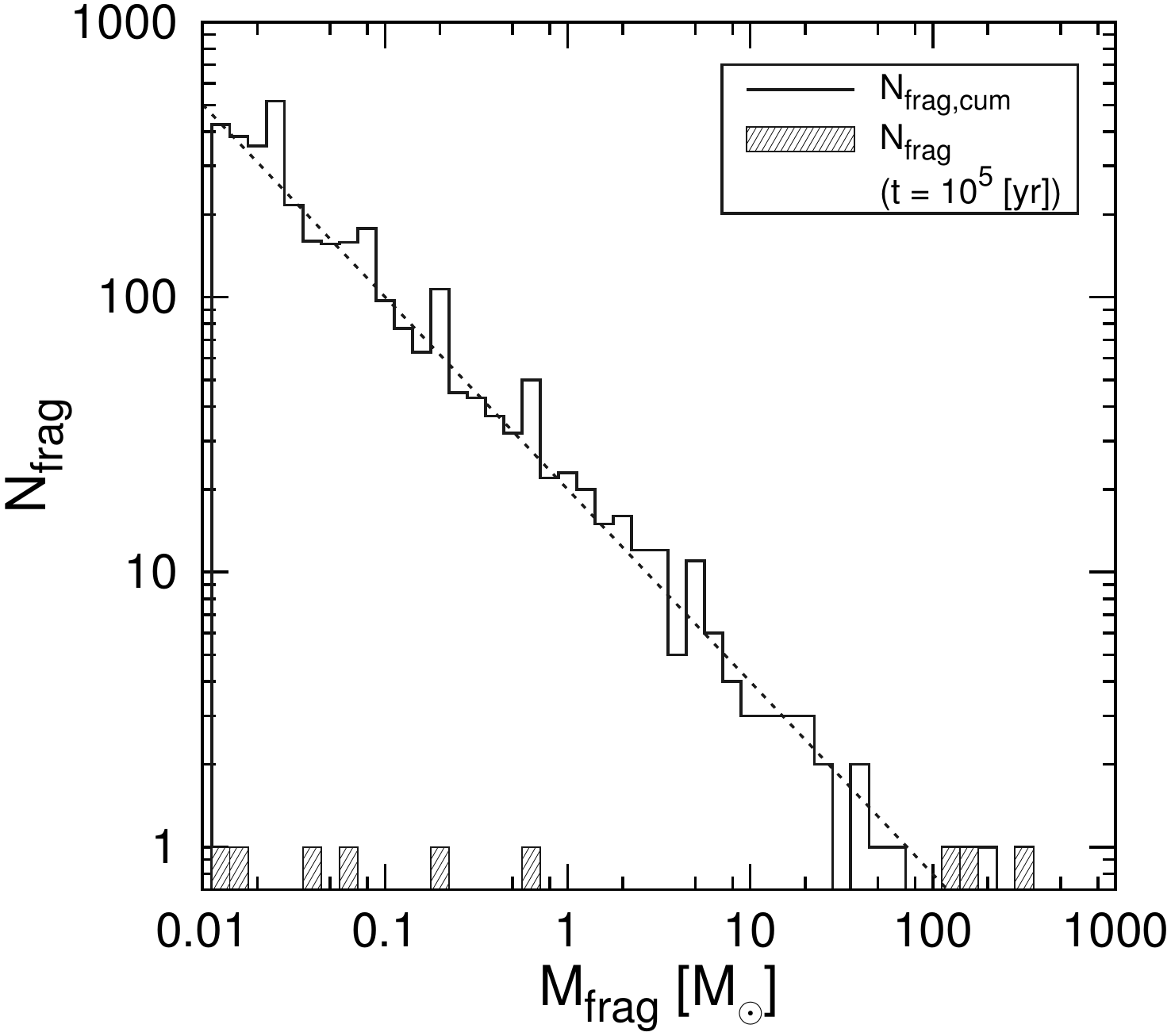}
\caption{
Mass spectrum of fragments, $t = 10^5$~yr after primary protostar formation, calculated with the same model as in Fig.~\ref{fig:Time-N+dNdt}. 
The grey-shaded histogram shows the mass spectrum of surviving fragments, whereas the steps show the cumulative distribution of total fragment number formed.
The dashed line represents a mass spectrum of $dN/dm \propto m^{-0.7}$, for ease of comparison.
}
\label{fig:Mfrag-Nfrag}
\end{center}\end{figure}

Here, we model the final fate of fragments with a simple continuity approach, to evaluate their resulting mass spectrum. 
The key concept is the balance between formation and migration rates, evaluated for the evolving accretion disc.
The details of our methodology are given in the Appendix~\ref{app:MergerModel}.
Our model is highly idealized, but it is useful to qualitatively understand the complex time evolution and final shape of the Pop~III stellar mass spectrum.
In Fig.~\ref{fig:Time-N+dNdt}, we illustrate the build-up of the mass spectrum over $10^5$~yr after primary protostar formation, with the fiducial parameter setting as employed in Section~\ref{sec:dis_angmom}.
Evidently, the fragment migration rate quickly equals the formation rate due to efficient angular momentum redistribution, such that the number of surviving fragments approaches an asymptotic plateau value, here of order 10.
The migration timescale becomes longer for fragments formed at larger radii, where the Jeans mass is also higher. Such more distant and more massive fragments are, therefore, likely to survive to the end.
Fig.~\ref{fig:Mfrag-Nfrag} shows the final fragment mass spectrum, $10^5$~yr after the calculation begins.
Fragments can form at every scale and with any mass, well fitted by $dN/dm \propto m^{-0.7}$, but most of them migrate to the disc centre.
The mass spectrum of surviving fragments exhibits a flat distribution,  $dN/dm \propto m^0$.
As a conclusion from this analysis, the efficient viscosity in the primordial gas disc causes only a limited number of fragments to survive, whose mass spectrum is nearly flat.

\subsection{Close BH binaries?} \label{sec:dis_closebinary}

We have shown that the fate of a fragment depends on the disc properties,
and its initial location therein.
The mass spectrum in Fig.~\ref{fig:Mfrag-Nfrag} implies that fragments with up to about $100~\msun$ can form and migrate to the disc centre, until stellar radiation feedback ends dissipative migration. 
Within this idealized picture, the most massive fragments form at the final stage of accretion, such that their migration timescale is long enough for them to survive (see Fig.~\ref{fig:Mfrag-Nfrag}).
It may thus be possible that some massive fragments migrate, but do not merge, such that close, massive binaries can form. 
In terms of the overall gas reservoir, there is sufficient fuel to form Pop~III stars which leave massive BHs behind, with dozens of solar masses each.
Such close binary of massive Pop~III BH remnants is one of the candidate sources for the gravitational wave signal recently observed. 
Specifically, Advanced LIGO has detected the signal from the coalescence of two black holes with $36^{+5}_{-4}$ and $29^{+4}_{-4}~\msun$ \citep[GW150914;][]{GW150914}.\footnote{The second detection indicates somewhat lower masses: $14.2^{+8.3}_{-3.7}$ and $7.5^{+2.3}_{-2.3}~\msun$ \citep[GW151226;][]{GW151226}.}
Interestingly, recent studies indicate that the GW signal may be helpful to restrict the IMF of Pop~III stars \citep{kinugawa16a,hartwig16}.
Gravitational wave astronomy may thus be able to open a new window into the cosmic dawn.

\section{Conclusions} \label{sec:sum}

To understand the final mass distribution of the first stars, one of the key remaining issues is the long-term evolution of individual protostars, whether they survive or merge.
To address this question, we perform a set of hydrodynamical simulations, without inserting sink particles. 
We instead use an opaque core model which enables us to examine the evolution of the protostellar accretion process over long durations, while considering the transport of fragments via viscous and dynamical angular momentum transfer.

We find that fragments formed anywhere in the accretion disc efficiently migrate to the centre within a local free-fall timescale, which is of the same order as the viscous timescale. 
The prevalence of rapid migration is caused by the high gas temperature in the primordial accretion disc, which in turn drives up the disc viscosity, $\nu_{\rm vis} = \alpha \cs H$.
Even if the disc contains metals and dust, as in the Pop~I case, the disc temperature declines, but the density also declines, such that the resulting viscous timescale becomes shorter than the free-fall time, for the fiducial viscous efficiency parameter.
In both primordial and present-day discs, almost all fragments can thus migrate to the primary protostar, until the central star evacuates the surrounding gas by its radiation.
If they were to finally merge, there would remain only a limited number of surviving fragments, and their mass spectrum would approach a flat distribution.

We have already gained considerable insight into the building process of the Pop~III IMF \citep[e.g.][]{clark11b,greif11,turk12,hosokawa11,stacy12,susa13}.
One important lesson is that the final IMF will be significantly affected by the efficiency of merging.
We therefore, ultimately, need to resolve the realistic hydrodynamics of the final merger process, because the sink technique cannot properly model this, and tends to underestimate the occurrence of mergers.
On the other hand, the alternative approach of using a stiffened equation of state to support a hydrostatic core, has problems as well. 
Since such cores are often artificially ``puffed up'', this technique tends to underestimate the dynamical effects of close encounters, and thus the resulting escape fraction of fragments.
Therefore, to robustly determine the final IMF of the first stars remains an open challenge.
It requires to simultaneously resolve the protostellar core at $\nh \sim 10^{22}~\cc$, and to continue the calculation until the ZAMS is reached, about $10^5$~yr after initial protostar formation. 
This, then, defines one of the frontiers in computational astrophysics, to be tackled over the coming years.

\section*{Acknowledgments}
We thank Takashi Hosokawa, Kohei Inayoshi, Tomoya Kinugawa, and Chon Summyon for helpful discussions.
We would also like to thank Paul C. Clark for his careful reading of the manuscript and valuable comments.
The numerical calculations were carried out on the Cray XC30 at the Center for Computational Astrophysics (CfCA) of the National Astronomical Observatory of Japan, the XC40 at YITP in Kyoto University, the Texas Advanced Computing Center (TACC) at the University of Texas at Austin, and COMA at the Center for Computational Sciences, University of Tsukuba.
This work was financially supported Grant-in-Aid for JSPS Overseas Research Fellowships (SH) and NSF grant AST-1413501 (VB).

\bibliographystyle{mnras}
\bibliography{biblio}

\appendix

\section{Population I disc model}
\label{app:PopIdisc}

\begin{figure}\begin{center}
\includegraphics[width=0.9\columnwidth]{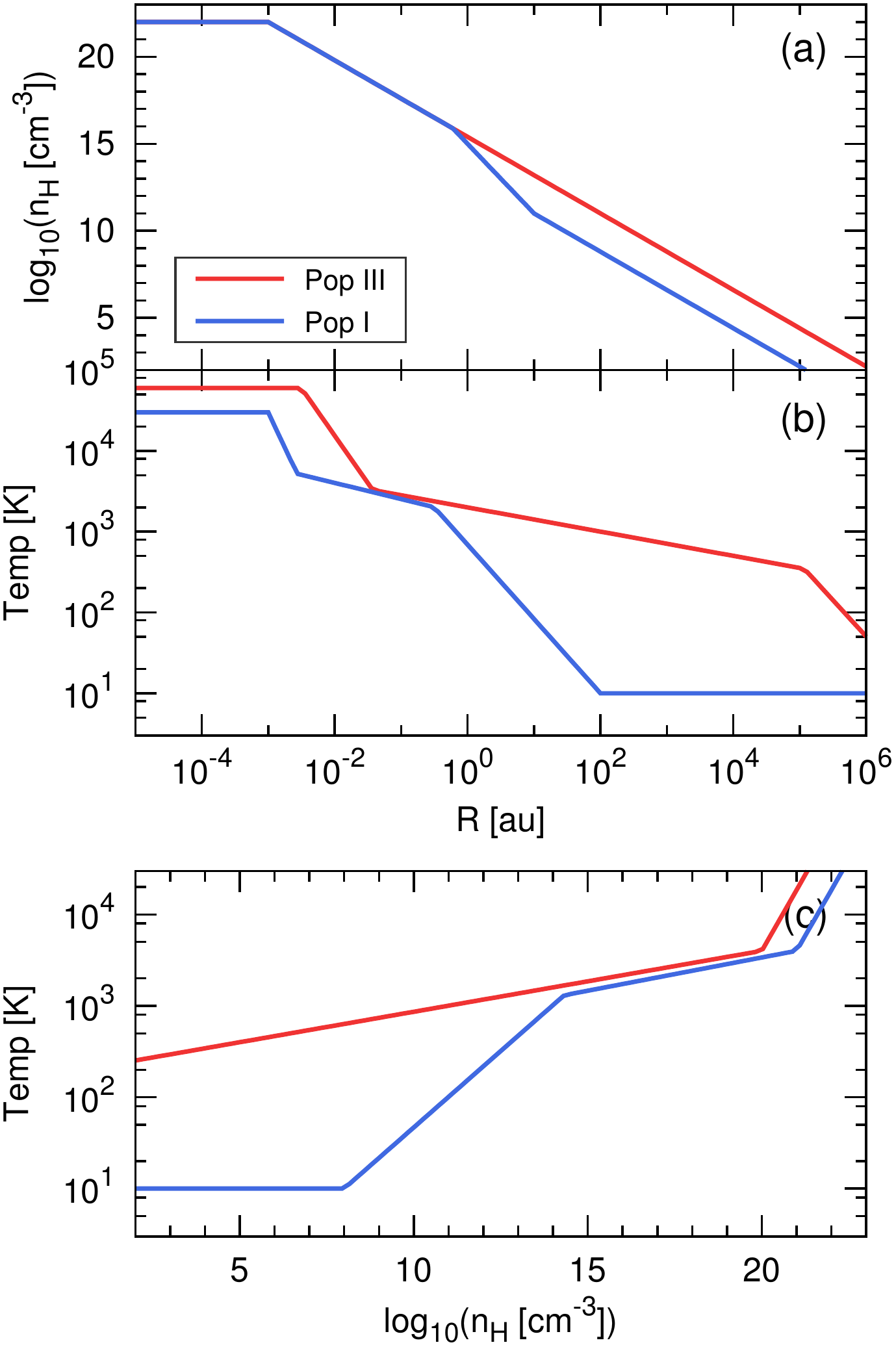}
\caption{
Density (panel a) and temperature (b) distributions of the star-forming clouds at the end of the cloud collapse for the primordial case \citep[Pop~III; fitted by fig.~7 in][]{omukai10} and the solar-metallicity case (Pop~I; fitted by fig.~13).
Panel (c) shows the temperature profile as a function of the gas number density.
The Pop~III profiles are well reproduced with the model adopted in the current study.
}
\label{figA:DiscModel}
\end{center}\end{figure}

To investigate the migration timescale in the present-day case, we adopt the density and temperature distributions of the collapsing clouds with solar metallicity, shown in fig.~13 in \cite{omukai10}.
Fig.~\ref{figA:DiscModel} summarizes these properties, and also shows the Pop~III case (fitted by fig.~7).
The fitted functions are
\begin{eqnarray}
n_{\rm H,I}(R) =
\begin{cases}
10^{22}~\cc \\
\ \ \ \ \ \ \ ({\rm for}~R/{\rm au} \le 10^{-3})~,\\
10^{22} (R/10^{-3}~{\rm au})^{-11/5}~\cc \\
\ \ \ \ \ \ \ ({\rm for}~10^{-3} < R/{\rm au} \le 0.6)~,\\
7.7 \times 10^{15} (R/0.6~{\rm au})^{-4}~\cc \\
\ \ \ \ \ \ \ ({\rm for}~0.6 < R/{\rm au} \le 10)~,\\
10^{11} (R/10~{\rm au})^{-11/5}~\cc \\
\ \ \ \ \ \ \ ({\rm for}~10 < R/{\rm au})~,\\
\end{cases}
\label{eqA:PopI_dens}
\end{eqnarray}
and
\begin{eqnarray}
T_{\rm I}(R) =
\begin{cases}
3.0 \times 10^4~{\rm K} \\
\ \ \ \ \ \ \ ({\rm for}~R/{\rm au} \le 10^{-3})~,\\
3.0 \times 10^4 (R/10^{-3}~{\rm au})^{-7/4}~{\rm K} \\
\ \ \ \ \ \ \ ({\rm for}~10^{-3} < R/{\rm au} \le 10^{-2.57})~,\\
3.3 \times 10^3 (R/10^{-2.57}~{\rm au})^{-1/5}~{\rm K} \\
\ \ \ \ \ \ \ ({\rm for}~10^{-2.57} < R/{\rm au} \le 10^{-0.51})~,\\
2.0 \times 10^3 (R/10^{-0.51}~{\rm au})^{-23/25}~{\rm K} \\
\ \ \ \ \ \ \ ({\rm for}~10^{-0.51} < R/{\rm au} \le 10^2) \\
10~{\rm K} \\
\ \ \ \ \ \ \ ({\rm for}~10^2 < R/{\rm au})~,\\
\end{cases}
\label{eqA:PopI_temp}
\end{eqnarray}
for the Pop~I case, whereas
\begin{eqnarray}
n_{\rm H,III}(R) =
\begin{cases}
10^{22}~\cc \\
\ \ \ \ \ \ \ ({\rm for}~R/{\rm au} \le 10^{-3})~,\\
10^{22} (R/10^{-3}~{\rm au})^{-11/5}~\cc \\
\ \ \ \ \ \ \ ({\rm for}~10^{-3} < R/{\rm au})~.\\
\end{cases}
\label{eqA:PopIII_dens}
\end{eqnarray}
and
\begin{eqnarray}
T_{\rm III}(R) =
\begin{cases}
6 \times 10^4~{\rm K} \\
\ \ \ \ \ \ \ ({\rm for}~R/{\rm au} \le 10^{-2.5})~,\\
6 \times 10^4 (R/10^{-2.5}~{\rm au})^{-6/5}~{\rm K} \\
\ \ \ \ \ \ \ ({\rm for}~10^{-2.5} < R/{\rm au} \le 10^{-1.45})~,\\
3.3 \times 10^3 (R/10^{-1.45}~{\rm au})^{-3/20}~{\rm K} \\
\ \ \ \ \ \ \ ({\rm for}~10^{-1.45} < R/{\rm au} \le 10^{5.04})~,\\
3.4 \times 10^2 (R/10^{5.04}~{\rm au})^{-9/10}~{\rm K} \\
\ \ \ \ \ \ \ ({\rm for}~10^{5.04} < R/{\rm au})~,\\
\end{cases}
\label{eqA:PopIII_temp}
\end{eqnarray}
for the Pop~III case.

We also notice that the adopted models represent the physical properties at the end of the cloud collapse.
In reality, the density and temperature distributions change during the accretion phase \citep[similar time evolution occurs in the Pop~III case; see fig.~11 in][]{clark11a}.
However, any such more detailed, time-resolved, analysis is beyond the scope of the current work, to be addressed in future work.

\section{Fragment merger history}
\label{app:MergerModel}

We here briefly discuss our idealized methodology to model the build-up of the final stellar mass spectrum (i.e. the IMF).
We solve for the time evolution of fragment number, as a function of mass, and of disc mass, as a function of the radial coordinate.
As initial conditions, we select a radial disc mass profile, $M_{\rm i}$ ($i = 1$ to $N$), mapped from the power-law density distribution in Equ.~\ref{eq:Menc-Rfrag}, with $A = 2.5 \times 10^{15}~\cc$, $b = 2.2$.
We compute gas number density as
\begin{eqnarray}
n_{\rm H,i} = \frac{M_{\rm i}}{\mh V_{\rm i}}~{\rm , where}~V_{\rm i} = \frac{4 \pi}{3} (r_{\rm i}^3 - r_{\rm i-1}^3)~.
\label{eqA:MeshDens}
\end{eqnarray}
We model the gas temperature, depending on gas density, with a fit to a pre-computed thermal evolution calculation as
\begin{eqnarray}
T_{\rm i} =
\begin{cases}
200~{\rm K}~(n_{\rm H,i} / 10^4~\cc)^{-0.233}\\
\ \ \ \ \ \ \ ({\rm for}~n_{\rm H,i}/\cc<10^4)~,\\
200~{\rm K}~(n_{\rm H,i} / 10^{4}~\cc)^{0.0822}\\
\ \ \ \ \ \ \ ({\rm for}~10^4<n_{\rm H,i}/\cc<10^{21})~,\\
5000~{\rm K}~(n_{\rm H,i} / 10^{21}~\cc)^{0.5}\\
\ \ \ \ \ \ \ ({\rm for}~n_{\rm H,i}/\cc>10^{21})~.\\
\end{cases}
\label{eqA:TempModel}
\end{eqnarray}
We update the free-fall timescale (Equ.~\ref{eq:t_ff}) and viscous timescale (Equ.~36), assuming $\alpha = 1.0$, and $\cs = 2.5~\kms$, calculated from local disc properties, at each timestep.
We again adopt the maximum of the free-fall and viscous timescales as migration timescale, $t_{\rm mig,i} = \max(t_{\rm ff,i}, t_{\rm vis,i})$, and define the mass accretion rate as $\dot{M}_{\rm i} = M_{\rm J,i} / t_{\rm mig,i}$.
With these assumptions, we calculate the evolution of disc mass as
\begin{eqnarray}
M_{\rm i}(t + dt) = M_{\rm i}(t) + \dot{M}_{\rm i} \cdot dt~.
\label{eqA:MassGrowth}
\end{eqnarray}

We define the following criteria for fragment formation: (1) the Jeans mass $M_{\rm J,i}$ (Equ.~\ref{eq:Mjeans}) exceeds $0.01~\msun$, which is the initial mass of a protostellar core; (2) the disc contains sufficient mass to become gravitationally unstable, such that $M_{\rm i} \le M_{\rm J,i}$; and (3) a free-fall time has passed after the previous fragmentation at the same location.
If these conditions are satisfied, a fragment with mass $M_{\rm J,i}$ forms, and its mass is subtracted from the local disc.
The fragment finally migrates, on the timescale discussed in Section~\ref{sec:dis_angmom}, and we remove it from the mass spectrum of surviving fragments.

\begin{figure}\begin{center}
\includegraphics[width=0.95\columnwidth]{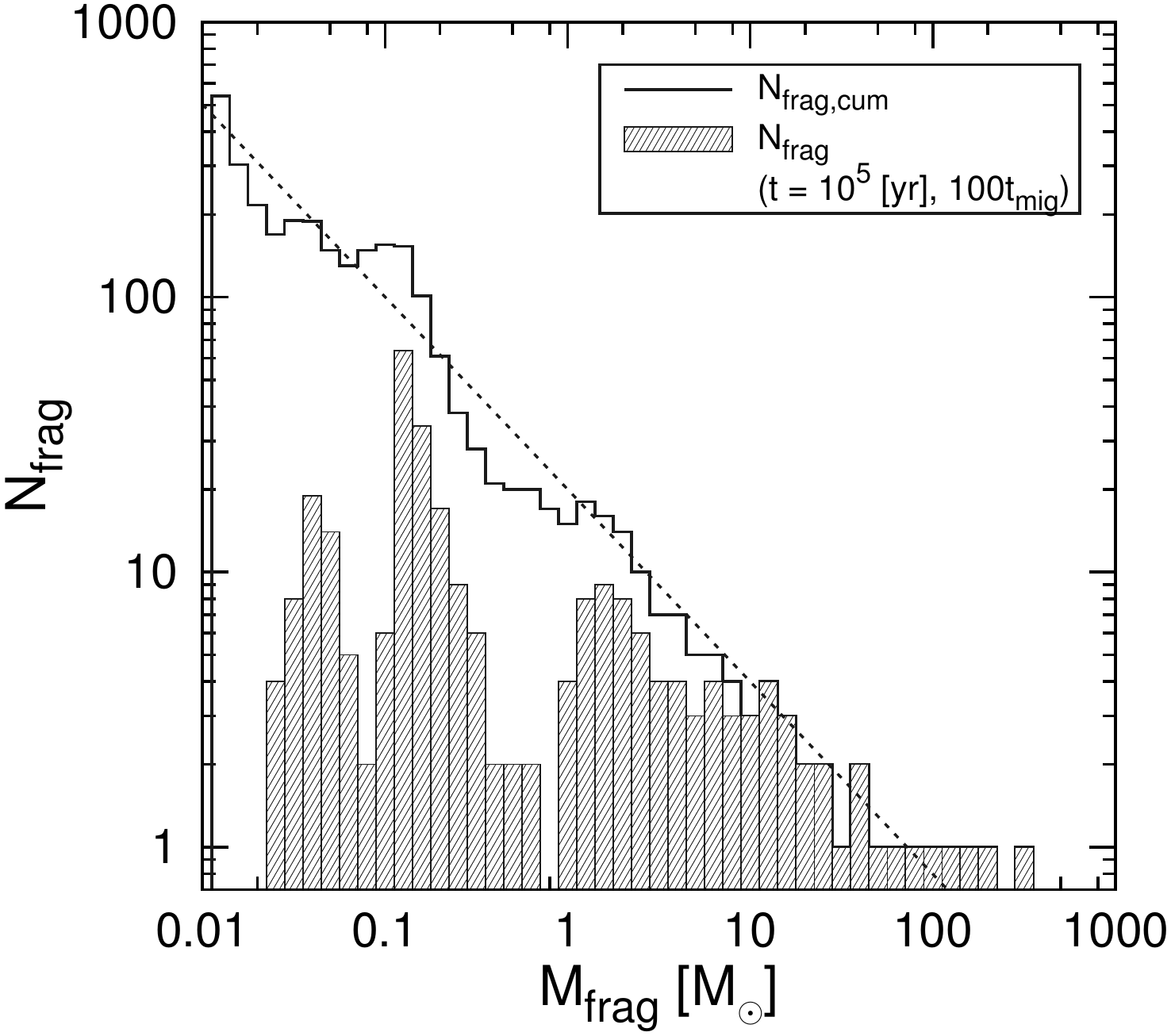}
\caption{
Pop~III mass function for inefficient migration. 
We here adopt the same conventions as in
Fig.~\ref{fig:Mfrag-Nfrag}, but use a much longer migration timescale, $t_{\rm mig} = 100 \max (t_{\rm ff}, t_{\rm vis})$.
The dashed line represents a power-law fit, where $dN/dm \propto m^{-0.7}$.
As can be seen, the resulting spectrum is now much closer to the unprocessed one at time of formation.
}
\label{figA:Mfrag-Nfrag_100times}
\end{center}\end{figure}

\begin{figure}\begin{center}
\includegraphics[width=0.95\columnwidth]{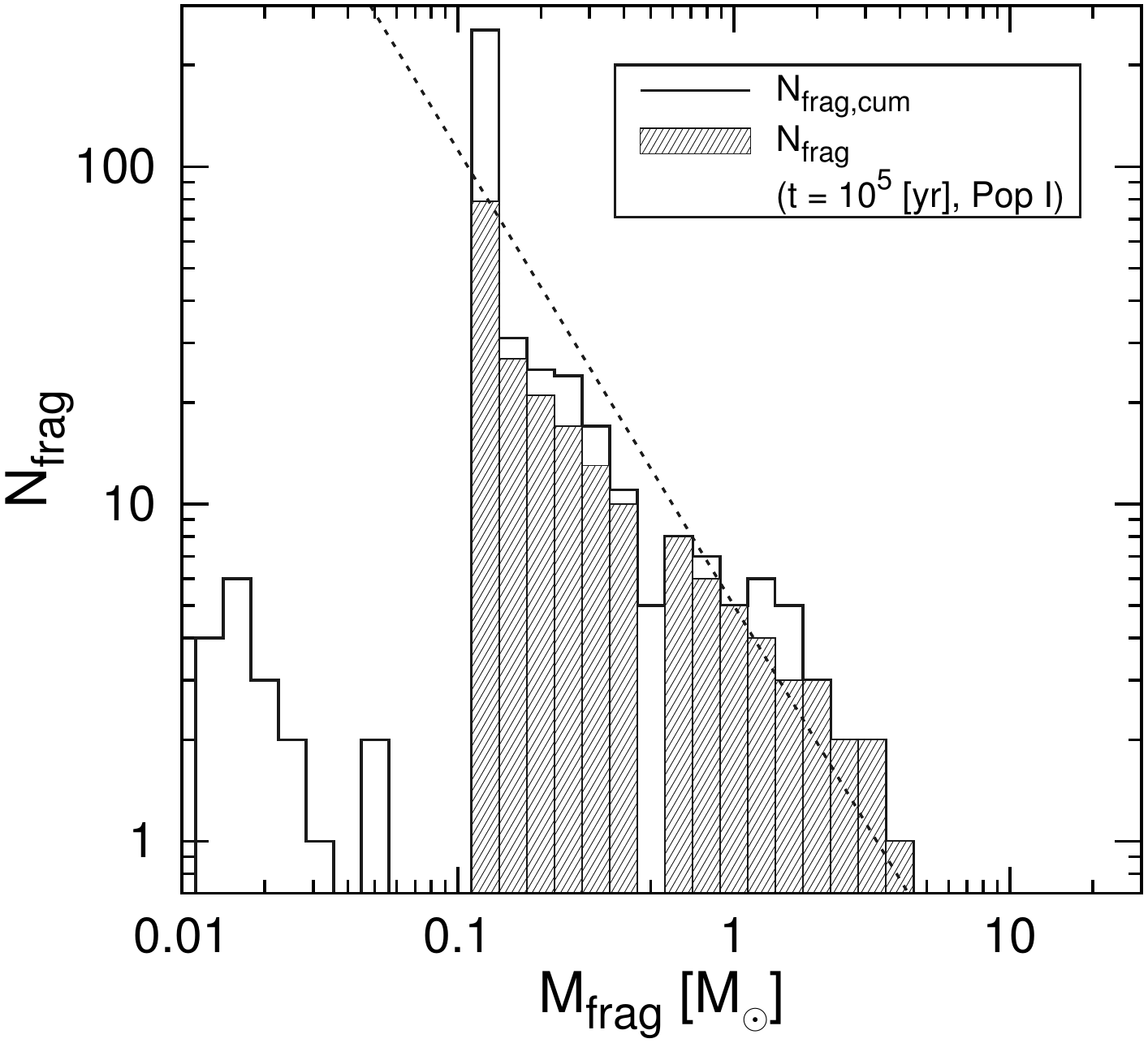}
\caption{
Pop~I mass function.
We again adopt the conventions of Fig.~\ref{fig:Mfrag-Nfrag}, but for the Pop~I disc with $\alpha = 1$.
The dashed line shows a fit to the mass spectrum, here $dN/dm \propto m^{-1.35}$.
Again, more low-mass stars survive, compared to the fiducial Pop~III model, thus reproducing the bottom-heavy
nature of the Pop~I IMF.
}
\label{figA:Mfrag-Nfrag_PopI}
\end{center}\end{figure}

The purpose of this toy model is to qualitatively evaluate the final fragment mass spectrum.
There are several processes, neglected here, which can affect the final results. Among them are dynamical escape due to N-body interactions, and tidal disruption.
On the other hand, we do model the broad features of fragments merging within a migration time, $t_{\rm mig,i}$.
In Fig.~\ref{figA:Mfrag-Nfrag_100times}, we show the resulting fragment mass spectrum in a similar fashion to Fig.~\ref{fig:Mfrag-Nfrag}, but now calculated with a migration time that is a 100 times longer.
In this setting, the formation rate becomes higher than the migration one, such that a considerable fraction of fragments can survive.
We can easily modify our model for the Pop~I case, employing the settings in Section~\ref{sec:dis_PopI}.
With the adopted density distribution for the Pop~I case, the total number of formed fragments is only about 10, significantly lower than that for the Pop~III case, due to the limited mass supply.
As one alternative condition, Fig.~\ref{figA:Mfrag-Nfrag_PopI} shows the resulting mass distribution formed on more massive Pop~I disc with adopting the Pop~III density distribution.
As discussed in Section~\ref{sec:dis_PopI}, the migration timescale is larger in a Pop~I disc, compared to Pop~III, such that the number of surviving fragments increases
In addition, the maximum fragment mass which can form and migrate within the stellar contraction time dramatically decreases.
Furthermore, the slope of the distribution steepens, from $\propto m^{-0.7}$ for Pop~III to $\propto m^{-1.35}$ for Pop~I, due to the smaller Jeans mass for colder discs.

\bsp
\label{lastpage}
\end{document}